\documentclass[sigconf, nonacm]{acmart}
\AtBeginDocument{%
  }

\renewcommand\footnotetextcopyrightpermission[1]{}
\acmConference{}{}{}

\usepackage{amsmath} 
\usepackage{balance} 
\usepackage{booktabs} 
\usepackage{caption}
\usepackage{wrapfig}
\usepackage{comment}
\usepackage{enumitem}
\setlist[itemize]{leftmargin=2.2em}

\usepackage{fancyvrb} 
\usepackage[symbol]{footmisc}
\usepackage{graphicx} 
\usepackage{makecell} 
\usepackage{multirow} 
\usepackage{soul}
\usepackage{subcaption}
\usepackage{xspace} 
\usepackage{bm}
\usepackage{cancel}
\usepackage[linesnumbered,ruled,vlined]{algorithm2e}
\usepackage{pifont}
\usepackage{xcolor} 
\usepackage{multicol}
\usepackage{xstring}

\usepackage{chngcntr} 

\theoremstyle{definition}

\theoremstyle{plain}

\theoremstyle{remark}

\newcommand\dbname{\ensuremath{\textsf{NeurCC}}\xspace}
\newcommand\polyjuice{{\text{Polyjuice}}}

\newcommand{\ignore}[1]{}

\newcommand\cacheF{\ensuremath{\mathcal{{F}}}}
\newcommand\modiV{\ensuremath{\mathcal{{V}}}}
\newcommand\cacheFT{\ensuremath{\mathcal{{F}}_{T}}}
\newcommand\cacheFP{\ensuremath{\mathcal{{F}}_{P}}}
\newcommand\cacheFD{\ensuremath{\mathcal{{F}}_{D}}}
\newcommand\encoder{\ensuremath{\mathcal{{E}}}}
\newcommand\inputstate{\ensuremath{\mathbf{{s}}}}
\newcommand\action{\ensuremath{\mathbf{{a}}}}
\newcommand\allfeatures{\ensuremath{\mathbf{{x}}}}

\definecolor{darkred}{RGB}{102, 0, 0}
\definecolor{darkgreen}{RGB}{0, 102, 0}

\newcommand\mypar[1]{%
  \par\addvspace{0.5\baselineskip}
  \noindent\textit{\textbf{#1}}.\ %
}
\makeatother

\usepackage{tabularx}
\newcolumntype{L}[1]{>{\raggedright\let\newline\\\arraybackslash\hspace{0pt}}m{#1}}

\newcommand{\comments}[2]{\noindent\underline{\textbf{#1}}: \emph{#2}}
\newcommand{\revisionpoint}[2]{
\par\medskip
\noindent\textbf{Reply for #1}: {#2}}

\newcommand{\revhighlight}[1]{\emph{\textbf{\uuline{#1}}}}
\newcommand{\question}[1]{{\textcolor{blue}{(#1)}}}
\newcommand{\edited}[1]{{#1}}

\colorlet{revR1}{darkgreen!60!black}
\colorlet{revR2}{red!70!black}
\colorlet{revR3}{blue!70!black}
\colorlet{revF}{red!70!black}
\colorlet{revReport}{orange!80!black}
\colorlet{revDefault}{black!70}

\newcommand{\setrevcolor}[1]{%
  \def\rev@color{revDefault}%
  \IfBeginWith{#1}{R1}{\def\rev@color{revR1}}{}%
  \IfBeginWith{#1}{R2}{\def\rev@color{revR2}}{}%
  \IfBeginWith{#1}{R3}{\def\rev@color{revR3}}{}%
  \IfBeginWith{#1}{F}{\def\rev@color{revF}}{}%
  \IfBeginWith{#1}{report}{\def\rev@color{revReport}}{}%
}

\long\def\revisionhint#1#2{%
  \begingroup
  \setrevcolor{#1}%
  \marginpar{%
    \textbf{\textcolor{\rev@color}{\raggedright\small #1}}%
  }%
  {\color{\rev@color}#2}%
  \endgroup
}

\renewcommand{\revisionhint}[2]{{\color{black}#2}}

\newif\ifextended\extendedtrue
\newcommand{\maintext}[1]{\ifextended\relax\else#1\fi}
\newcommand{\extended}[1]{\ifextended#1\else\relax\fi}
\setcopyright{none}

\begin{document}

\maintext{
\title{Modeling Concurrency Control as a Learnable Function}
}

\extended{
\title{Modeling Concurrency Control as a Learnable Function}}

\author{
  Hexiang Pan$^1$,
  Shaofeng Cai$^1$,
  Tien Tuan Anh Dinh$^2$,
  Yuncheng Wu$^3$, \\
  Yeow Meng Chee$^4$,
    Gang Chen$^5$,
  Beng Chin Ooi$^5$}

\affiliation{%
 \institution{$^1$National University of Singapore, $^2$Deakin University, $^3$Renmin University of China $^2$,   \\
$^4$Singapore University of Technology and Design, $^5$Zhejiang University}
\textit{panh@u.nus.edu}, \quad \textit{shaofeng@comp.nus.edu.sg}, \quad \textit{anh.dinh@deakin.edu.au}, \quad \textit{wuyuncheng@ruc.edu.cn}, \\  \textit{ymchee@sutd.edu.sg}, \quad \textit{\{cg, ooibc\}@zju.edu.cn}
}

\renewcommand{\shortauthors}{Pan et al.}

\renewcommand{\comments}[2]{#2}                 
\renewcommand{\revisionpoint}[2]{}             
\renewcommand{\revhighlight}[1]{#1}            
\renewcommand{\question}[1]{}                  
\renewcommand{\edited}[1]{#1}                  
\long\def\revisionhint#1#2{#2}
\colorlet{revR1}{black}
\colorlet{revR2}{black}
\colorlet{revR3}{black}
\colorlet{revF}{black}
\colorlet{revReport}{black}
\colorlet{revDefault}{black}

\begin{abstract}
Concurrency control (CC) algorithms are important in modern transactional databases, as they enable high performance by executing transactions concurrently while ensuring correctness. However, state-of-the-art CC algorithms struggle to perform well across diverse workloads, and most do not consider workload drifts.

In this paper, we propose NeurCC, a novel learned concurrency control algorithm that achieves high performance across diverse workloads. The algorithm is quick to optimize, making it robust against dynamic workloads. It learns a function that captures a large number of design choices from existing CC algorithms. The function is implemented as an efficient in-database lookup table that maps database states to concurrency control actions. The learning process is based on a combination of Bayesian optimization and a novel graph reduction search algorithm, which converges quickly to a function that achieves high transaction throughput. We compare NeurCC against five state-of-the-art CC algorithms and show that it consistently outperforms the baselines both in transaction throughput and in optimization time.

\end{abstract}

\maketitle

\section{Introduction}
\label{sec:introduction}

Concurrency control (CC) algorithms allow database transactions to exploit hardware concurrency while ensuring
correctness, i.e., transaction isolation. As modern database systems are no longer disk-bound, CC is
essential for achieving high transaction processing performance.
Classical CC algorithms, such as two-phase locking (2PL) and optimistic concurrency control (OCC), do not perform well across all workloads~\cite{DBLP:journals/pvldb/YuBPDS14/1000_core_cc, DBLP:conf/cidr/TangJE17/ACC}.
In particular, 2PL typically outperforms OCC for high-contention workloads, but the reverse is true for low-contention workloads.

Recent works on CC combine multiple classical CC algorithms to support different workloads.
They adopt the {\em classify-then-assign} approach, in which transaction operations are classified into different types, then the most appropriate CC algorithm is selected for each type.
The classification can be based on the transaction~\cite{DBLP:conf/mascots/TaiM96/AEP,DBLP:conf/icde/Su0Z21/C3,
DBLP:conf/sigmod/SuCDAX17/Tebaldi,DBLP:conf/sosp/XieSLAK015/Callas}, or on the data~\cite{DBLP:conf/sigmod/ShangLYZC16/HSync, DBLP:conf/cidr/TangJE17/ACC,DBLP:conf/icde/LinCZ18/ASOCC,DBLP:conf/usenix/TangE18/CormCC, DBLP:conf/osdi/WangDWCW0021/polyjuice}.
The selection of CC algorithm is either heuristic~\cite{DBLP:conf/icde/Su0Z21/C3,     DBLP:conf/mascots/TaiM96/AEP, DBLP:conf/icde/LinCZ18/ASOCC, DBLP:conf/usenix/TangE18/CormCC, DBLP:conf/sigmod/SuCDAX17/Tebaldi, DBLP:journals/pvldb/HongZLDCPZ25/HDCC} or based on learning~\cite{DBLP:conf/cidr/TangJE17/ACC, DBLP:conf/osdi/WangDWCW0021/polyjuice}.
We note that these works do not perform well for dynamic workloads that can drift over time. In particular, they require manual, time-consuming changes to the classification and algorithm selection when there are drifts.
CormCC~\cite{DBLP:conf/usenix/TangE18/CormCC} and ACC~\cite{DBLP:conf/cidr/TangJE17/ACC} propose dynamic mechanisms that reduce the cost of manual changes, but they only support a small number of CC algorithms.
{\polyjuice}~\cite{DBLP:conf/osdi/WangDWCW0021/polyjuice} uses an evolutionary algorithm to learn the optimal combination of CC algorithms.
It learns {\em CC actions} --- e.g., to wait for dependent transactions to reach specific steps, or to expose uncommitted operations --- for each data access.
However, it does not support general transactions and does not work well for dynamic workloads due to the long optimization time.

Our goal is to design a learned CC algorithm that can achieve high performance across diverse workloads, and is
efficient to optimize. More specifically, the algorithm can extract a high degree of concurrency from different
workloads, while incurring small overhead. 
The short optimization time allows it to perform well under
dynamic workloads.
There are three challenges in designing this algorithm. 
First, 
a large number of CC algorithms proposed in the literature
are optimized for specific
workloads. 
Thus, for our algorithm to perform well across different workloads, it must incorporate design choices from as
many existing CC algorithms as possible.
Second, to achieve high performance, the algorithm's overhead must be  smaller than the transaction execution time.
This is non-trivial because, in modern in-memory databases, transactions can be executed in microseconds (or even in
nanoseconds).
Third, learning the optimal action for each transaction operation is challenging because most of the learned parameters
(such as the wait parameters in {\polyjuice}) are non-continuous. This lack of locality renders machine learning techniques such as Policy Gradient~\cite{DBLP:conf/nips/SuttonMSM99}
ineffective, thus algorithms like Polyjuice resort to costly optimization strategies such as evolutionary algorithms and random search.

\begin{figure}
  \centering
\includegraphics[width=0.85\linewidth]{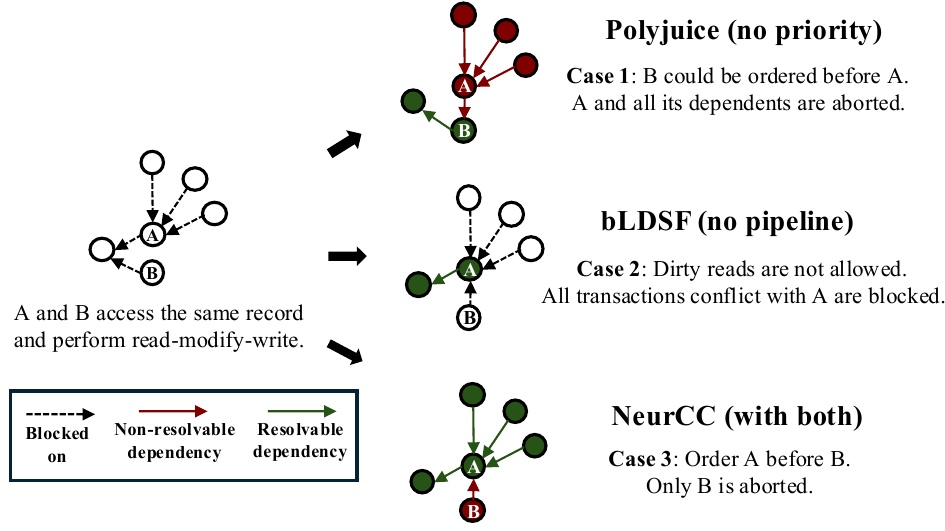}
    \vspace{-1.5em}
  \caption{{Example of CC designs covered by different algorithms.}}
    \vspace{-1em}
  \Description{Diagram showing concurrency control designs not covered by the Polyjuice system.}
  \label{fig:polyjuice-uncover}
\end{figure}

We present {\dbname}, a novel learned CC algorithm that achieves the goal above. It addresses the first challenge by
proposing a unified model of existing concurrency control algorithms. In particular, {\dbname} models a concurrency control
algorithm as a learnable function \({\cacheF}: \{{\inputstate}\} \to \{{\action}\}\) that maps the database state
{\inputstate} to a set of concurrency control actions {\action}. It captures the existing \textit{classify-then-assign} approaches:
{\inputstate} represents the features used for classification, {\action} indicates the actions of the assigned CC algorithm.
In {\dbname}, a state {\inputstate} consists of features used by existing adaptive concurrency controls, such as data
hotness~\cite{DBLP:conf/icde/LinCZ18/ASOCC} and the number of dependent transactions in the dependency graph~\cite{DBLP:journals/pvldb/TianHMS18/LDSF}. An action
{\action} consists of a combination of conflict detection and resolution mechanisms.
Instead of selecting from existing algorithms, \dbname decomposes concurrency control into a sequence of conflict detection and resolution actions, which allows for a wide range of CC designs by enabling arbitrary combinations of these actions.
The model is more general than existing adaptive concurrency control algorithms, such as {\polyjuice}~\cite{DBLP:conf/osdi/WangDWCW0021/polyjuice} and bLDSF~\cite{DBLP:journals/pvldb/TianHMS18/LDSF}.
The former only maps data access IDs to pipeline-wait actions, and the latter only maps dependency set sizes to priority-related wait actions. 
\revisionhint{R2.M1}{Figure~\ref{fig:polyjuice-uncover} illustrates the CC designs covered by {\polyjuice} and bLDSF. Specifically, {\polyjuice} treats all transactions accessing the same data as equals, which could result in suboptimal actions with high abort rates.
bLDSF gives higher priority to transactions with more dependent transactions, but it does not consider transaction pipelining, which is essential for efficient transaction execution. 
}

{\dbname} addresses the second challenge by making the collection of the state {\inputstate} and the execution of  {\cacheF}({\inputstate}) highly efficient. Specifically, it excludes states that are expensive to compute, e.g., hash
values~\cite{DBLP:conf/icde/Su0Z21/C3} and conflict graphs~\cite{DBLP:conf/sigmod/WangMCYCL16/ic3}, and instead uses states
that can be computed in a lock-free manner, such as the number of direct dependencies and the number of executed
operations. {\dbname} implements {\cacheF} as an in-database lookup table for the state-action pairs.
This table only consumes a small amount of memory, as {\dbname} learns an efficient feature selector that reduces the input states to only hundreds of
values.
The whole table fits in the database cache, and the CC actions for an operation can be determined by a single table lookup,
which only costs hundreds of CPU cycles.

{\dbname} addresses the third challenge in three steps. First, it reduces the cost of evaluating the function's performance.
Previous works evaluate the function's performance by loading the lookup table into the database and monitoring the system performance.
This process takes seconds to minutes~\cite{DBLP:conf/osdi/WangDWCW0021/polyjuice}. {\dbname} adopts a Bayesian
optimization approach that trains a surrogate model for predicting the system performance given an action lookup
table. By sampling from this surrogate model, it can quickly explore the search space and guide the optimization
process to select the next function to evaluate.
Second, {\dbname} re-formulates the original learning task of pipeline-wait actions as an optimal
conflict graph learning task that exhibits locality and therefore is amenable to common learning techniques.
This reformulation is important since learning the pipeline-wait actions directly is challenging due to the lack of locality around the
optimal combination of them. In particular, a small modification to an action could lead to dependency cycles and
significant performance reduction. {\dbname} uses an efficient algorithm similar to genetic search for this conflict graph learning task.
Third, {\dbname} speeds up the learning process by 
prioritizing the actions that yield higher improvements.
Specifically, it prioritizes fine-tuning wait-related actions that can bring higher performance improvements over actions like timeouts and back-off durations.

We compare {\dbname} against state-of-the-art CC algorithms, including the recent learned CC algorithms~\cite{DBLP:conf/usenix/TangE18/CormCC, DBLP:conf/osdi/WangDWCW0021/polyjuice}.
The results demonstrate that
{\dbname} consistently outperforms the baselines across diverse workloads.
{
The algorithm achieves up to 3.32${\times}$ higher throughput and 11${\times}$ faster optimization time than the state-of-the-art learned baseline, Polyjuice~\cite{DBLP:conf/osdi/WangDWCW0021/polyjuice}, making it more robust against workload drifts.
}

In summary, we make the following contributions:

\begin{itemize}[leftmargin=*]

\item We present {\dbname}, a novel learned concurrency control algorithm that achieves high performance across diverse
workloads. The algorithm is quick to optimize, allowing it to adapt quickly to workload drifts.

\item We propose a novel model of concurrency control, which is a function mapping database states into
conflict-handling actions. This model captures a large number of existing CC algorithms.

\item We design an efficient learning algorithm that quickly finds a high-performance concurrency control model.

\item 
{We conduct extensive evaluation of {\dbname}, comparing it against five state-of-the-art CC algorithms,
{namely 2PL~\cite{DBLP:journals/tods/RosenkrantzSL78/2PL_wait_die}, Silo~\cite{DBLP:conf/sosp/TuZKLM13/silo}, CormCC~\cite{DBLP:conf/usenix/TangE18/CormCC}, Polyjuice~\cite{DBLP:conf/osdi/WangDWCW0021/polyjuice}, and IC3~\cite{DBLP:conf/sigmod/WangMCYCL16/ic3}.}
The results show that {\dbname} achieves high throughput and low optimization time.
Its throughput is up to 3.32${\times}$, 4.38${\times}$, and 4.27${\times}$
higher than that of {\polyjuice}, 2PL, and Silo, respectively. 
}
%
\end{itemize}

The remainder of the paper is structured as follows.
Section~\ref{sec:background} discusses the design space of concurrency control algorithms.
%
Section~\ref{sec:overview} provides an overview of {\dbname}. Section~\ref{sec:cc_design} details how {\dbname} performs
fine-grained concurrency control via an optimized action lookup table. Section~\ref{sec:fine-tuning} describes the
learning process. %
Section~\ref{sec:deployment} outlines the
deployment workflow. %
Section~\ref{sec:evaluation} presents the experimental evaluation.
Section~\ref{sec:related} discusses related works, and Section~\ref{sec:conclusion} concludes.

\section{Concurrency Control Design Space} \label{sec:background}

In this section, we first describe the design space of concurrency control (CC) algorithms. We then discuss how
existing algorithms fit into the space, and identify the gaps representing new opportunities.

A transaction is modeled as a sequence of data access operations, each being either a read or a write.  CC algorithms
allow transactions to be executed concurrently and correctly~\cite{DBLP:conf/sigmod/AdyaGLM95/occ,
DBLP:conf/sosp/TuZKLM13/silo, DBLP:reference/db/Fekete18c/ssi, DBLP:conf/sigmod/LimKA17/cicada,
DBLP:conf/sigmod/WangMCYCL16/ic3, DBLP:journals/pvldb/WangK16/mocc, DBLP:conf/sigmod/GuoWYY21/bamboo,
DBLP:conf/eurosys/MuAS19/DRP, DBLP:journals/pvldb/FaleiroAH17/EWV, 2PL_no_wait,
DBLP:journals/tods/RosenkrantzSL78/2PL_wait_die, DBLP:journals/pvldb/BarthelsMTAH19/2PL-BW,
DBLP:journals/tods/HsuZ92/cautious_wait, DBLP:conf/icde/FranaszekRT91/wait_depth_limit,
DBLP:conf/sigmod/ShengTZP19/abort_prediction, DBLP:conf/pods/CareyKL90/half_half, DBLP:conf/vldb/MoenkebergW92/opt_DMP,
DBLP:conf/icde/MoenkebergW91/conflict_ratio, DBLP:conf/sigmod/ThomsonDWRSA12/Calvin,
DBLP:journals/pvldb/KallmanKNPRZJMSZHA08/HStore, DBLP:conf/sigmod/SuCDAX17/Tebaldi, DBLP:conf/usenix/TangE18/CormCC,
DBLP:conf/sosp/XieSLAK015/Callas, DBLP:conf/osdi/WangDWCW0021/polyjuice, DBLP:journals/pvldb/LuYCM20/Aria,
DBLP:conf/cidr/TangJE17/ACC, DBLP:conf/icde/LinCZ18/ASOCC, DBLP:conf/sigmod/ShangLYZC16/HSync,
DBLP:conf/icde/Su0Z21/C3, NeurDB}. To ensure correctness, a CC algorithm employs {\em CC actions} --- algorithmic steps that ensure
safe interleaving of data access operations among concurrent transactions. In the following, we describe the design space
of CC actions consisting of two design dimensions: conflict detection and conflict resolution. A CC algorithm can use
one or multiple actions from this space.

\begin{figure}
  \centering
\includegraphics[width=0.98\linewidth]{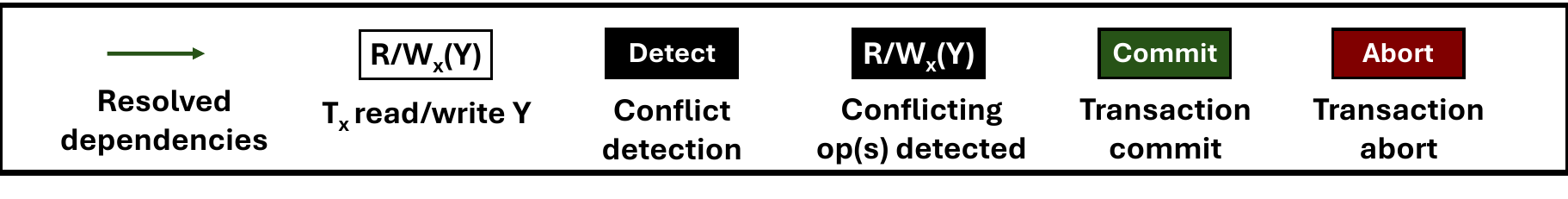}
\captionsetup[subfigure]{font=scriptsize, justification=centering}
\begin{subfigure}{0.23\linewidth}
\includegraphics[width=\linewidth]{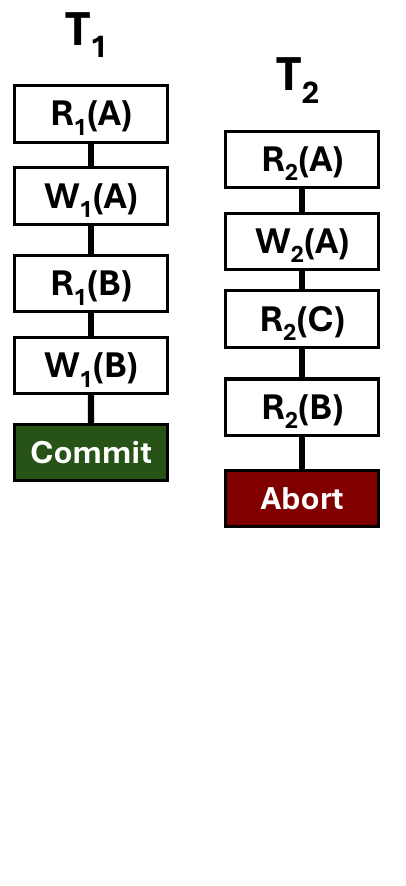}
\subcaption{No conflict detection}
\label{fig:no-detect}
\end{subfigure}
\hfill
\vline%
\hfill
\begin{subfigure}{0.23\linewidth}
\includegraphics[width=\linewidth]{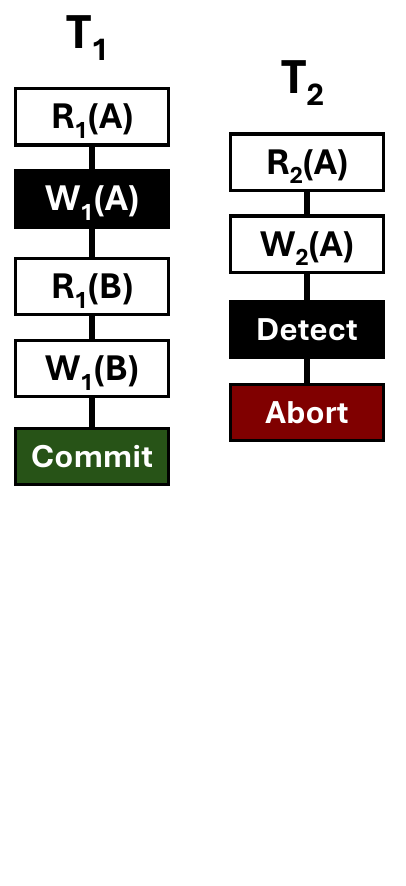}
\subcaption{Detect critical conflicts (abort)}
\label{fig:detect-critical-abort}
\end{subfigure}
\hfill
\vline%
\hfill
\begin{subfigure}{0.23\linewidth}
\includegraphics[width=\linewidth]{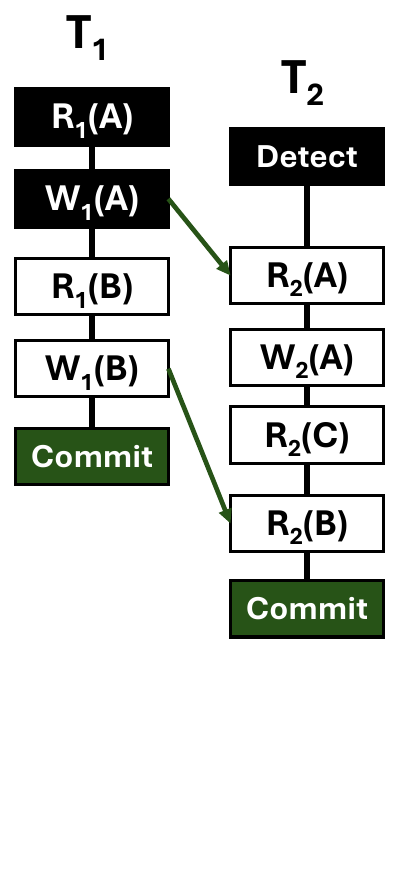}
\subcaption{Detect critical conflicts (block)}
\label{fig:detect-critical-block}
\end{subfigure}
\hfill
\vline%
\hfill
\begin{subfigure}{0.23\linewidth}
\includegraphics[width=\linewidth]{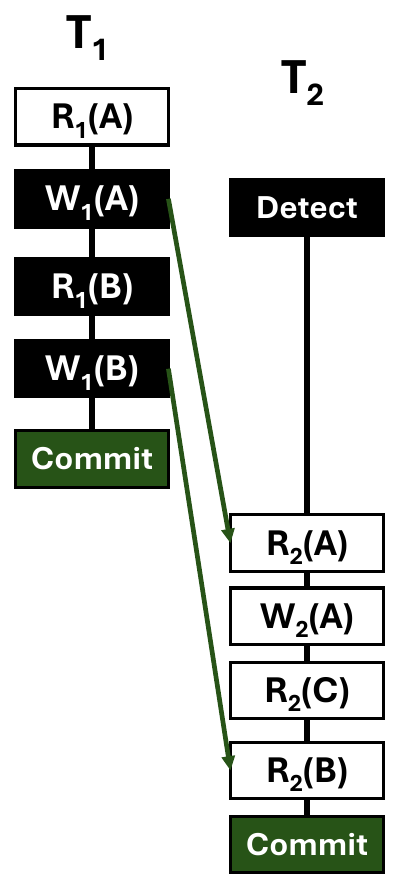}
\subcaption{Detect all conflicts}
\label{fig:detect-all}
\end{subfigure}
\vspace{-3.5mm}
\caption{Examples of different conflict detection approaches. $T_2$ has RW conflicts with $T_1$ on A and B.}
\vspace{-1em}
\Description{Diagram showing different conflict detection approaches.}
\label{fig:cd}
\end{figure}

\subsection{Conflict Detection}
\label{sec:background.conflict_detection}
A conflict occurs when two operations from two concurrent transactions access the same data object, and one of the
accesses is a write. Existing CC algorithms use different approaches to detect conflicts, based on their assumptions about the
distribution and probability of conflicts occurring during transaction execution. We identify three design choices for
conflict detection: no detection, detecting critical conflicts, and detecting all conflicts. The first option
optimistically assumes that there are no conflicts and therefore implements no conflict detection. The third option
pessimistically assumes many conflicts and therefore detects them as soon as possible to avoid costly aborts. The second option is in between: it detects some critical conflicts while ignoring others.

We consider two transactions $T_1$ and $T_2$ executing concurrently. They both perform read/write on data object A
and B. Figure~\ref{fig:cd} illustrates examples where $T_2$ has Read-Write (RW) conflicts with $T_1$, and $T_2$ uses
different conflict detection approaches. 

\mypar{No detection}
The CC algorithm assumes conflicts are rare and performs no conflict detection during execution. 
It relies on validation at commit time for correctness. The optimistic concurrency control (OCC) and its variants follow this
approach. In particular, each transaction keeps its writes locally (as opposed to exposing them to other transactions),
and assumes no other concurrent transactions will update its read data or read its uncommitted writes. The algorithm allows transaction execution to proceed without the overhead of conflict detection but incurs overhead (in aborts) when there are conflicts.  Figure~\ref{fig:no-detect} shows an example where $T_2$ aborts during the final validation because $T_1$
updated its previously read A and B.

\mypar{Detect critical}
The CC algorithm avoids the cost of aborts by identifying conflicts that are likely to cause aborts based on heuristic rules. They are called
{\em critical} conflicts, which may lead to validation failure or to dependency cycles. Upon detection, the algorithm
immediately aborts the transaction {(early validation)}, e.g., ASOCC~\cite{DBLP:conf/icde/LinCZ18/ASOCC} and
SSI~\cite{DBLP:reference/db/Fekete18c/ssi}, or reorders operations to allow conflicting transactions to commit together {(pipeline-wait)},
e.g., Bamboo~\cite{DBLP:conf/sigmod/GuoWYY21/bamboo} and IC3~\cite{DBLP:conf/sigmod/WangMCYCL16/ic3}.
In Figure~\ref{fig:detect-critical-abort}, $T_2$ detects that $T_1$ updated the previously read A during early
validation and immediately aborts without executing to the end.
In Figure~\ref{fig:detect-critical-block}, $T_2$ detects that the immediate read of A would form a dependency cycle
between $T_1$ and $T_2$. It then reorders its read of A to occur after $T_1$'s write of A, thus breaking the cycle.

\mypar{Detect all}
The CC algorithm detects all conflicts that may lead to transaction aborts. The classical 2PL and its
variants~\cite{2PL_no_wait,  DBLP:journals/tods/RosenkrantzSL78/2PL_wait_die, DBLP:journals/pvldb/BarthelsMTAH19/2PL-BW} adopt this approach.
Figure~\ref{fig:detect-all} shows an example in which this approach causes $T_2$ to block for a long time, which
increases transaction latency. Furthermore, the detection algorithm introduces some overhead because it has to find all
conflicts. 


\subsection{Conflict Resolution}
\label{sec:background.conflict_resolving}
Once a conflict is detected, the concurrency control algorithm resolves it by coordinating accesses from the conflicting
transactions. We identify two design choices for conflict resolution: timeout and priority.

\mypar{Timeout}
When the CC algorithm detects a conflict, it blocks the current transaction for a specific duration (a timeout), waiting for the conflicting operations to end in this duration, and aborts it when the timeout is exceeded.
A timeout of $0$ means that the transaction aborts immediately~\cite{2PL_no_wait}, which assumes pessimistically that the transaction will eventually abort.
A large timeout, or timeout of $\infty$, means the transaction is blocked.
In particular, the transaction is either put on a wait list associated with the data object, or is entering a spinning loop until the data object is available.
This approach is implemented in lock-based and pipeline-wait-based CC algorithms, e.g., 2PL and IC3. The blocking overhead can be significant, especially when the transaction eventually aborts.

\mypar{Priority}
When multiple transactions are blocked on the same data object, the algorithm may order them based on their priorities.
When the object becomes available, the highest-priority transaction is woken up first.
{Most existing CC algorithms either do not consider priority~\cite{DBLP:conf/sosp/TuZKLM13/silo, DBLP:journals/tods/HsuZ92/cautious_wait, DBLP:conf/osdi/WangDWCW0021/polyjuice}, or use simple FIFO or timestamp priority~\cite{DBLP:journals/pvldb/BarthelsMTAH19/2PL-BW, DBLP:journals/tods/RosenkrantzSL78/2PL_wait_die, DBLP:conf/sigmod/GuoWYY21/bamboo, DBLP:journals/pvldb/ZhangXWYLGL25/Rebirth-Retire}.
Recent works make heuristic priority assignment to conflicting transactions~\cite{DBLP:conf/eurosys/HuangMW17/VProfiler, DBLP:conf/sigmod/HuangMSW17/VATS, DBLP:journals/pvldb/TianHMS18/LDSF, DBLP:journals/pacmmod/YeHCY23/Polaris}.}
For example, bLDSF~\cite{DBLP:journals/pvldb/TianHMS18/LDSF} gives higher priorities to transactions with higher numbers of dependent transactions.  Other algorithms~\cite{DBLP:conf/eurosys/HuangMW17/VProfiler, DBLP:conf/sigmod/HuangMSW17/VATS} consider transaction age and the number of held locks when assigning priorities.

\begin{table*}[htbp]
    \centering
    \caption{{Comparison of state-of-the-art concurrency control algorithms and \dbname in the design space. Timeout is omitted because it is supported by all algorithms. {Tuned priority refers to either heuristic or learned priority assignments.}}}
    \vspace{-3mm}
    \renewcommand{\arraystretch}{1.0}
    \setlength{\tabcolsep}{6pt}
    \begin{tabular}{c|c|c|c|c|c}
    \specialrule{.1em}{.05em}{.05em} 
    \multirow{2}{*}{} & & \multicolumn{2}{c|}{Resolve w/o tuned priority} & \multicolumn{2}{c}{Resolve with tuned priority} \\
    \cline{2-6}
       & {No detection} & {Detect critical}  &  {Detect all} & {Detect critical} &  {Detect all} \\
    \specialrule{.1em}{.05em}{.05em}
       OCC~\cite{DBLP:conf/sigmod/AdyaGLM95/occ}, Silo~\cite{DBLP:conf/sosp/TuZKLM13/silo}   & \checkmark & & & & \\
       \hline
       2PL~\cite{2PL_no_wait}, TSO~\cite{DBLP:books/aw/BernsteinHG87/tso1, DBLP:conf/icde/LeisK014/tso2}, MVCC~\cite{wu2017empirical/mvcc} & & & \checkmark & & \\ 
       \hline
       Bamboo~\cite{DBLP:conf/sigmod/GuoWYY21/bamboo}, Rebirth-Retire~\cite{DBLP:journals/pvldb/ZhangXWYLGL25/Rebirth-Retire}, EWV~\cite{DBLP:journals/pvldb/FaleiroAH17/EWV}, IC3~\cite{DBLP:conf/sigmod/WangMCYCL16/ic3} & & \checkmark & & & \\ 
       \hline
       bLDSF~\cite{DBLP:journals/pvldb/TianHMS18/LDSF}, VATS~\cite{DBLP:conf/sigmod/HuangMSW17/VATS} & & & & & \checkmark \\ 
       \hline
       Polaris~\cite{DBLP:journals/pacmmod/YeHCY23/Polaris}  & \checkmark & & & & \checkmark \\ 
       \hline
        MCC~\cite{DBLP:conf/sosp/XieSLAK015/Callas}, Tebaldi~\cite{DBLP:conf/sigmod/SuCDAX17/Tebaldi} & & \checkmark & \checkmark  & & \\ 
       \hline
       ACC~\cite{DBLP:conf/cidr/TangJE17/ACC}, TxnSails~\cite{DBLP:journals/pvldb/ZhuangLLCSZSPD25/TxnSails}, C3~\cite{DBLP:conf/icde/Su0Z21/C3}, CormCC~\cite{DBLP:conf/usenix/TangE18/CormCC}, HDCC~\cite{DBLP:journals/pvldb/HongZLDCPZ25/HDCC} & \checkmark & & \checkmark &  & \\ 
       \hline
       ASOCC~\cite{DBLP:conf/icde/LinCZ18/ASOCC}, {\polyjuice}~\cite{DBLP:conf/osdi/WangDWCW0021/polyjuice} & \checkmark & \checkmark & \checkmark & & \\ 
       \hline
       {\bf \dbname} & \checkmark & \checkmark & \checkmark &  \checkmark & \checkmark \\ 
    \specialrule{.1em}{.05em}{.05em}
    \end{tabular}
    \vspace{-1em}
    \label{tab:overview-existing}
\end{table*}

\subsection{State-of-the-Art Concurrency Control Algorithms}

Table~\ref{tab:overview-existing} compares state-of-the-art concurrency control algorithms within the design space. We omit timeout from the table because it is supported by all algorithms. 
It can be seen that some algorithms, such as OCC and 2PL, support only one type of CC actions.
Other algorithms, such as {\polyjuice}~\cite{DBLP:conf/osdi/WangDWCW0021/polyjuice} and ASOCC~\cite{DBLP:conf/icde/LinCZ18/ASOCC}, support multiple CC actions, with which the algorithm can fine-tune conflict handling actions based on the workload to extract more concurrency.

{The table reveals several gaps in the design space, i.e., CC actions that are not covered by existing algorithms. For example, no existing algorithms consider priority-based conflict resolution in combination with detecting critical conflicts. Most algorithms, including the ones that support multiple CC actions,  do not consider {tuned} transaction priorities.}

\section{\dbname Overview} \label{sec:overview}

In this section, we present an overview of {\dbname}, a novel learning-based concurrency control algorithm that achieves high performance across diverse workloads.
It supports the complete set of CC actions shown in Table~\ref{tab:overview-existing}, by capturing existing concurrency control algorithm designs via a learnable function. In the following, we describe the learning function, and the workflow that enables its efficient execution and optimization.
We summarize the notations in Table~\ref{tab:notation}.

\subsection{{Concurrency Control As a Learnable Function}}
\label{sec:cc_model}

\begin{table}
  \centering
  \caption{Summary of notations}
  \vspace{-3mm}
  \label{tab:notation}
  \begin{tabular}{ll}
    \toprule
    $T$ &   current transaction \\
    $op$ &  current operation of the current transaction \\
    ${\allfeatures}$ &  vector with all features \\
    ${\inputstate}$ &  database state \\
    ${\action}$ &  CC actions for the current operation \\
    ${\cacheF}$ &   function \\
    {${\encoder}$} & {feature selector} \\
    {Score({\cacheF})} & {system performance using {\cacheF}} \\
    \bottomrule
  \end{tabular}
\end{table}

{
{We model concurrency control algorithm as a function {\cacheF} that maps the current database state {\inputstate} to a combination of conflict detection and conflict resolution actions {\action}.}
%
%
\begin{align}
    \cacheF(\inputstate) = \action.
\label{eq:agent-function}
\end{align}
{The state {\inputstate} comprises features used in various CC algorithms, such as transaction access ID, transaction type, and data hotness. The actions {\action} include conflict detection actions, timeout values, and pipeline-wait actions.}
Before executing an operation, the algorithm collects the current database state {\inputstate}, computes \cacheF(\inputstate), and then executes the resulting actions {\action}.
%

{\cacheF} captures many existing CC algorithms, including {\polyjuice}, ASOCC, 2PL, bLDSF, OCC, and IC3.
We demonstrate this with two examples of 2PL (wait-die) and ASOCC.
%
Let $t_\textit{req}$ and $t_\textit{owner}$ be the timestamps of the current transaction and of the oldest lock owner.
Let $hotness$ be the hotness level of the data object being accessed.
2PL (wait-die) and ASOCC can be modeled as follows:


{
\[
    \mathcal{F}_{\text{2PL}}(t_{req}, t_{owner}) =\left\{
    \begin{aligned}
        (\text{\textit{detect all}}, 0, \varnothing) & & t_{req} > t_{owner} \\
        (\text{\textit{detect all}}, \infty, \textit{timestamp pri.}) & & \text{otherwise} \\
    \end{aligned}
    \right.
\]
\[
    \mathcal{F}_{\text{ASOCC}}(hotness) =\left\{
    \begin{aligned}
        (\text{\textit{no detection}}, 0, \varnothing) & & hotness = \text{cold} \\
        (\text{\textit{detect all}}, \infty, \varnothing) & & hotness = \text{hot} \\
        (\text{\textit{detect critical}}, 0, \varnothing) & & \text{otherwise} \\
    \end{aligned}
    \right.
\]
}

\noindent
Each action consists of a conflict detection action, a timeout value, and a transaction priority {($\varnothing$ represents no priority)}.
In these examples, $t_{req}$, $t_{owner}$, and $hotness$ are the state $\mathbf{s}$.
The output of $\mathcal{F}_{\text{2PL}}$ and $\mathcal{F}_{\text{ASOCC}}$ are then executed by the algorithm.
{In particular, 2PL (wait-die) aborts the transaction when $t_\textit{req}$ is higher than $t_\textit{owner}$; otherwise, it blocks the transaction and wakes it up later in the timestamp order.}
ASOCC ignores conflicts for cold data. It detects all conflicts and blocks the transaction for hot data.
In other cases, it aborts the transaction when there are critical conflicts.


{We define the problem of finding {\cacheF} that extracts the most concurrency across diverse workloads as an optimization problem.}
Let $\mathbf{S}$ be the set of possible input states {\inputstate}, Score(\cacheF) be the system performance\footnote{We use throughput as the performance metric, similar to previous works~\cite{DBLP:conf/osdi/WangDWCW0021/polyjuice, DBLP:conf/cidr/TangJE17/ACC}.} using {\cacheF}, and $\mathbf{A}$ be the set of possible output
actions {\action}.
The optimization objective is defined as:
\begin{align}
    \text{maximize} \quad & \text{Score}(\cacheF) \label{eq:offline-optimization} \\
    \text{subject to} \quad & \forall \inputstate \in \mathbf{S}, \ \action = \cacheF(\inputstate) \in \mathbf{A}. 
\end{align}
\revisionhint{R2.O7 F.Q2}{
{\dbname} starts the optimization process when it detects a workload drift.
Specifically, there is a background worker, named $optimizer$, that continuously monitors the system’s \emph{current} throughput, i.e. the current number of committed transactions per second (TPS). 
%
%
The $optimizer$ tracks the throughput regardless of whether the system is saturated.
Optimization 
is triggered when the relative change in throughput exceeds a predefined threshold.
In the current design, the threshold value 
is determined empirically, e.g., $10\%$.} 
The value is chosen to achieve reliable detection while avoiding unnecessary optimization triggers.
\maintext{We discuss this threshold in Section~\ref{sec:evaluation.stored}, and provide more details in  the extended version~\cite{NeurCC-extended}.}
\extended{We discuss this threshold in Section~\ref{sec:evaluation.stored}, and provide more details in Section~\ref{sec:func_update}.}
{We note that more complex workload drift detection techniques (e.g., two-sample-test~\cite{DBLP:journals/pacmmod/WuI24/2sample}) could also be used.
Since our main focus is on the learning of concurrency control, we leave the design and integration of advanced drift detectors to future work.}
%

\subsection{\dbname Workflow}
\label{sec:workflow}

\begin{figure*}[ht]
  \centering
  \includegraphics[width=\linewidth]{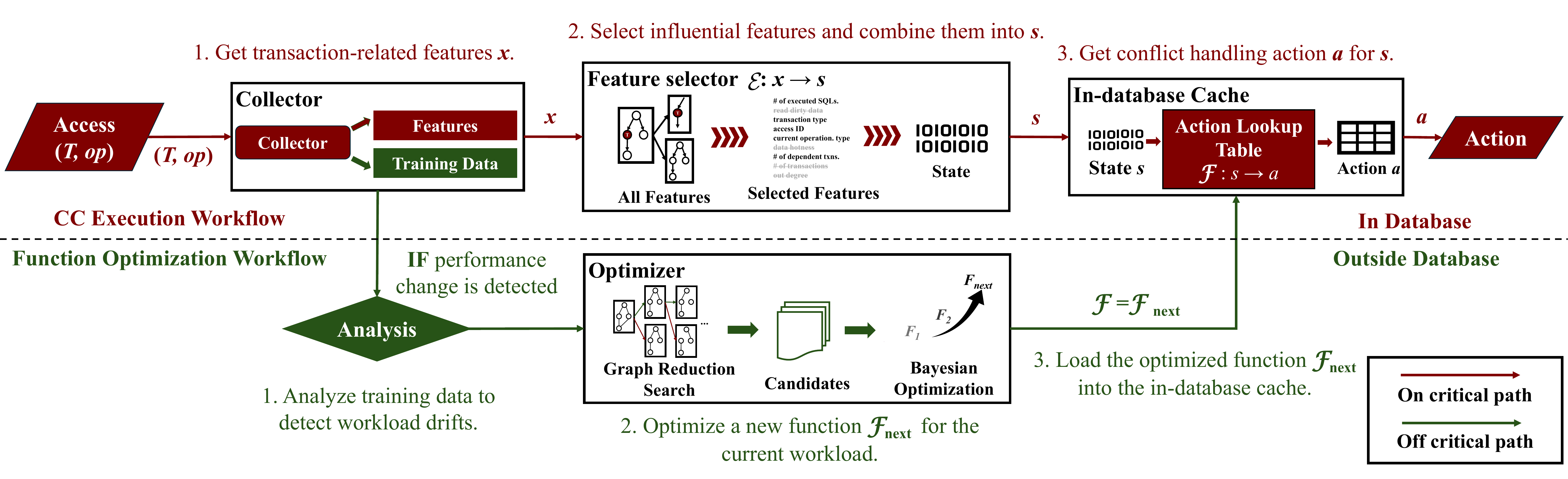}
  \vspace{-2.8em}
  \caption{{\dbname}'s concurrency control (CC) execution workflow and function optimization workflow.}
  \vspace{-1.6em}
  \Description{{\dbname}'s concurrency control (CC) execution workflow and function optimization workflow.}
  \label{fig:overview}
\end{figure*}

Figure~\ref{fig:overview} shows the main components of {\dbname}: a \textit{collector} that
tracks real-time database states, a \textit{feature selector} {\encoder} that maps the collected states into a state vector
{\inputstate}, an \textit{in-database cache} that maintains the function {\cacheF} in the form of an action lookup table, and
an \textit{optimizer} that detects workload drifts and optimizes the function upon detection.
%

%
{\dbname} operates across two workflows: the \textcolor{darkred}{\textit{CC execution workflow}} and the
\textcolor{darkgreen}{\textit{function optimization workflow}}.
The former is on the critical path of transaction execution. It is invoked before every data access operation to
select the optimal actions.
This workflow takes the current operation $op$ and transaction $T$ as input, and outputs a vector of actions {\action} for the transaction manager to execute. The transaction manager first sends $(T, op)$ to the collector, which
collects database states based on $T$ and $op$, then combines them into a feature vector {\allfeatures}.
Next, the feature selector $\encoder$ selects important features from {\allfeatures} and generates a state {\inputstate}.
Finally, the concurrency control actions {\action} are returned from the lookup table, i.e.,
$\action = \cacheF(\inputstate)$. 
This workflow is efficient such that it can be executed at per-operation granularity. We describe it in more detail in Section~\ref{sec:cc_design}.

The second workflow is off the critical path of transaction execution. It maintains high performance under workload drifts by monitoring the system performance and optimizing the action lookup table when necessary.
In this workflow, the function {\cacheF} and the corresponding system performance
\(\text{Score}(\cacheF)\) are collected periodically. The data is analyzed to detect workload drifts based on performance changes. When a drift is detected, {\dbname} starts optimizing the function for the current workload. The optimization process runs with a pre-defined time budget $t_\textit{max}$, at the end of which the new function 
$\mathcal{F}_\textit{next}$ is loaded into the database cache as a new action lookup table.
We describe this workflow in more detail in Section~\ref{sec:fine-tuning}.


\subsection{{Discussion}}
\label{sec:scope}

\revisionhint{R3.O1}{{\dbname} evaluates the agent function at per-operation granularity. This is because important CC mechanisms, such as dirty-read visibility and pipeline waits, work at the level of individual operations (as opposed to at the level of individual transaction). 
We note that coarse-grained decisions can be beneficial, for example, in enabling scheduling operations before execution (as in deterministic CC). Integrating such coarse-grained decisions with our fine-grained agent function is left as future work.}

\revisionhint{R3.O2}{The current design of {\dbname} targets CC paradigms based on per-operation conflict handling, which keeps the state and action space of the agent function small and efficient to optimize. We note that {\dbname} captures some transaction-chopping behaviors through its pipeline-wait actions (i.e., $w_i$ and $expose$), and it can reproduce the behavior of algorithms like IC3~\cite{DBLP:conf/sigmod/WangMCYCL16/ic3} that uses transaction chopping.
Other paradigms, such as deterministic CC, sagas, or timestamp-ordering, entail more complex state and action representations (e.g., actions that schedule transactions before execution).
Supporting them would significantly expand the  agent function’s state and action space, thus requiring new learning algorithms, which we leave for future work.}
%

\section{\dbname Execution}
\label{sec:cc_design}

In this section, we describe {\dbname}'s {execution workflow} which lies on the critical path of transaction execution.
We first discuss what database states {\dbname} collects, and how it selects important states as input to the function {\cacheF}. We then present the detailed concurrency control algorithm that performs conflict handling based on the output of {\cacheF}.

\subsection{{State Collection and Selection}}
\label{sec:capture_contention}

{Before executing a transaction operation, {\dbname} first computes a vector $\mathbf{s}$ that captures the current, relevant database states (Figure~\ref{fig:encoder}).}
{The vector {\inputstate} is important as it determines the quality of CC actions generated by {\cacheF}.}
{There are two challenges in computing {\inputstate}.}
First, it must be efficient since it is on the critical path of transaction execution. Modern in-memory databases can commit a transaction in microseconds, therefore {\inputstate} should be computed in order of nanoseconds.
Second, {\inputstate} must contain relevant database states so that CC actions returned by {\cacheF} can extract a high degree of concurrency, while remaining small to minimize storage overhead and to speed up the optimization of {\cacheF}.

\begin{table}[t]
\centering
\small
\maintext{
\caption{\textcolor{revR2}{Features collected by {\dbname}. $T$ denotes the current transaction.}}}
\extended{\caption{Features collected by {\dbname}. $T$ denotes the current transaction.}}
\label{tab:features}
{\vspace{-3mm}}
\renewcommand{\arraystretch}{1.0}
\setlength{\tabcolsep}{6pt}
\begin{tabular}{p{2.4cm} p{5.6cm}}
\toprule
\textbf{Feature} & \textbf{Definition} \\
\midrule
\# of executed SQLs
& Number of SQLs already executed by $T$. \\

Read dirty data
& If $T$ has read uncommitted dirty data from others. \\

Transaction type
& Identifier of the transaction template to which $T$ belongs (e.g., \texttt{NewOrder}, \texttt{Payment}). \\

Access ID
& Identifier of the currently executing SQL. \\

Operation type
& The current operation type (read/write). \\

Data hotness
& Contention level of the accessed record (\textit{cold}, \textit{warm}, \textit{hot}), derived from data access frequency and data correlation~\cite{DBLP:conf/icde/LinCZ18/ASOCC}. \\

\# of dependent txns.
& Number of transactions $T$ depends on. \\

\# of transactions
& Number of running transactions. \\

Out-degree
& Number of transactions that depend on $T$. \\
\bottomrule
\end{tabular}
\end{table}

{\dbname} employs two techniques to address the challenges above.
The first is to use a fixed set of lightweight states (or features) that can be efficiently captured across transactions, in a lock-free manner.
\dbname collects nine features, \revisionhint{R2.O5}{which are listed in Table~\ref{tab:features}} and can be grouped into transaction features and graph features.
These features have been shown to be effective at capturing transaction and dependency graph characteristics~\cite{DBLP:journals/pvldb/TianHMS18/LDSF, DBLP:conf/icde/LinCZ18/ASOCC, DBLP:conf/osdi/WangDWCW0021/polyjuice, DBLP:conf/sigmod/WangMCYCL16/ic3, DBLP:conf/sigmod/HuangMSW17/VATS, DBLP:conf/sigmod/SuCDAX17/Tebaldi}.
More importantly, they are common features readily available in transactional databases, that is, they are not specific to any concurrency control algorithm.
{The second technique is to use a feature selector {\encoder} that selects influential features to achieve a balance between the quality of the optimized function and the optimization time.}
{\dbname} provides a default {\encoder} that achieves good performance, which the user can improve at some cost, by running the optimization algorithm described in Section~\ref{sec:prune_opt}.

\begin{figure}[htbp]
  \centering
  \includegraphics[width=0.9\linewidth]{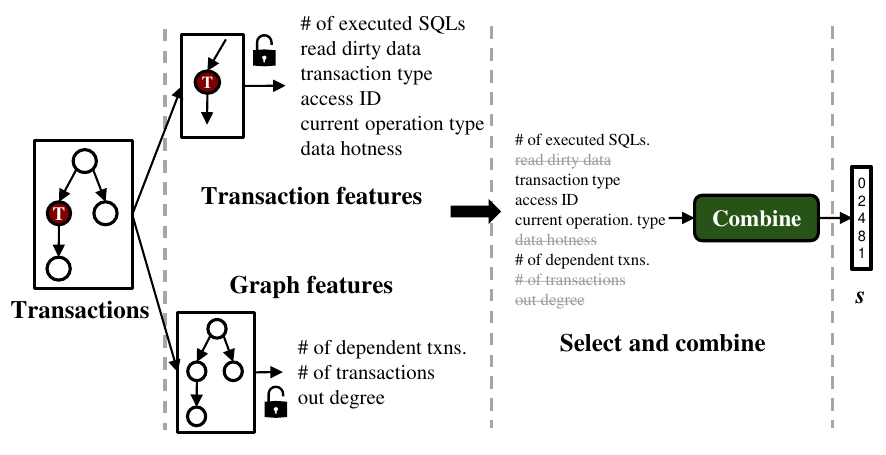}
  \vspace{-4mm}
  \caption{{{\dbname}'s state collection and selection process.}}
  \Description{{\dbname}'s state encoding process.}
  \label{fig:encoder}
  \vspace{-1mm}
\end{figure}

\revisionhint{R1.O1 R1.O2}{Before a transaction accesses a data record, {\dbname} collects the current database state as a vector {\inputstate}.
Consider ASOCC as a running example.
This vector includes data hotness feature, thus access to records with different contention levels is mapped to different states, which enables the selection of appropriate conflict-handling actions as in ASOCC.}

\subsection{{Action Function}}
\label{sec:get_action}

After computing {\inputstate}, {\dbname} executes \cacheF({\inputstate}) to get the concurrency control actions {\action} for the current transaction operation.
The main challenge in designing {\cacheF} is to ensure both efficiency and effectiveness.
Specifically, the function must be quick to evaluate, since it is on the transaction's critical path. Furthermore, the output actions must be able to 
extract high degrees of concurrency across different workloads.  For efficiency, {\dbname} implements {\cacheF} as a
learnable action lookup table, reducing the function evaluation to a single table lookup. Although implementing {\cacheF} as a machine learning model can approximate the state-action mapping more accurately, the inference cost is too high for concurrency control. In contrast, the lookup table can return  the actions in hundreds of CPU cycles. 

\begin{figure}[htbp]
  \centering
  \includegraphics[width=\linewidth]{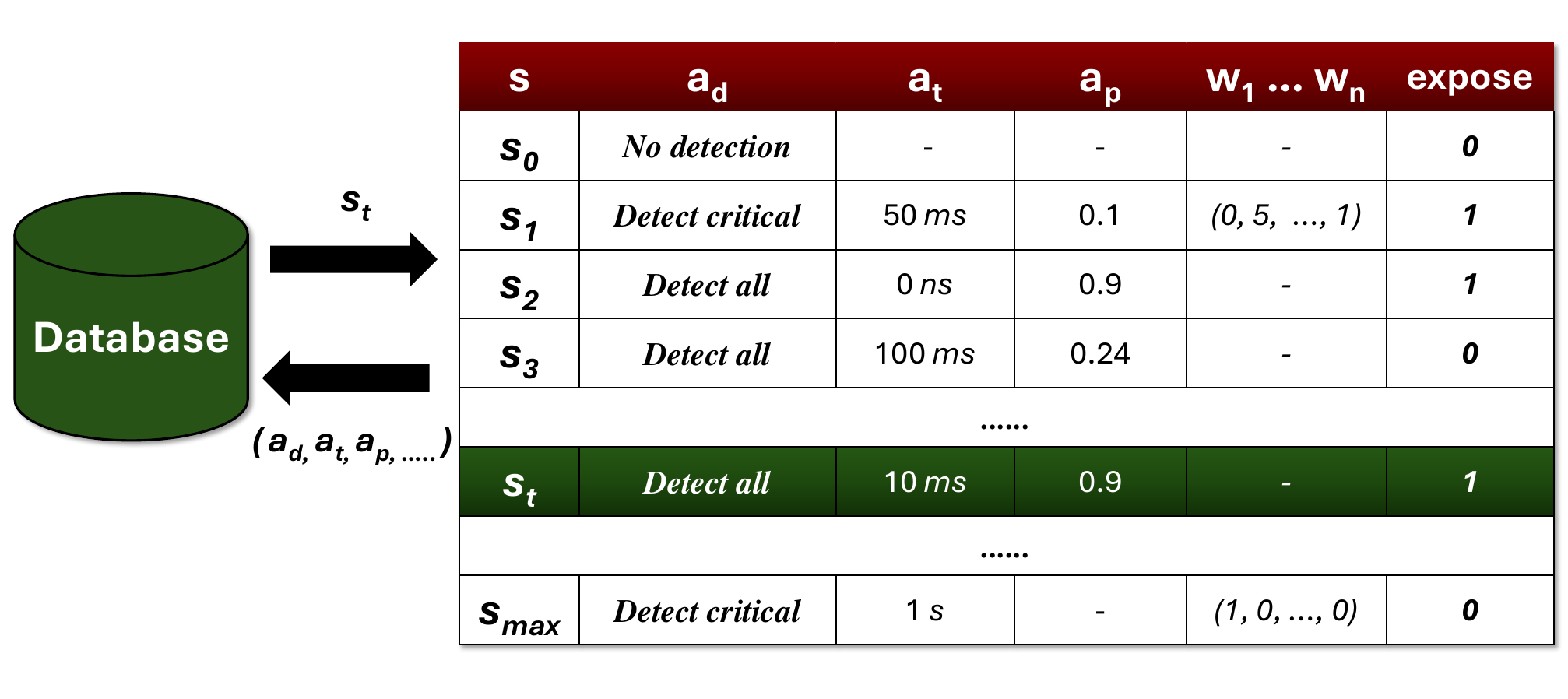}
  \vspace{-7mm}
  \caption{{{\dbname} function implemented as a lookup table}.}
  \Description{{\dbname}'s action lookup table.}
  \label{fig:policy}
  \vspace{-2mm}
\end{figure}

To achieve high performance across diverse workloads, {\cacheF} learns combinations of design choices from existing CC algorithms optimized for different workloads.
An output of {\cacheF} contains a conflict detection mechanism, which can be either \textit{no detection}, \textit{detect critical}, or \textit{detect all}.
It also contains a timeout value and a transaction priority for conflict resolution. 
Figure~\ref{fig:policy} shows an action lookup table, where the first column is the input state and the remaining columns
make up the CC actions.
Let $n$ denote the number of transaction types. The table has the following schema:
\begin{align}
\label{eq:cc-api}
& \langle \action_d, \action_t, \action_p, (w_1, \dots, w_n, \textit{expose}) \rangle \\
\text{where} \quad & \action_d \in \{ \text{\textit{no detection}}, \text{\textit{detect critical}}, \text{\textit{detect all}} \} \\
& \action_t \in [0, +\infty), \quad \action_p \in [0, 1] \\
& w_i \in \mathbb{N}, \quad expose \in \{0, 1\}
\end{align}

\noindent $\action_d$, $\action_t$, and $\action_p$ are the conflict detection method, timeout, and transaction priority, respectively.
$(w_1, \dots, w_n)$ represents the rule for identifying critical conflicts (as described in
Section~\ref{sec:background.conflict_detection}).
{
Specifically, $w_i$ means that $T$ considers the first $w_i$ accesses of a dependent transaction with type $i$ as critical conflicts, thereby preventing the formation of dependency cycles among concurrent transactions that might read dirty data from each other.
$expose$ determines if $T$ makes its accesses visible to other transactions after executing $op$.
}

\revisionhint{R1.O1 R1.O2}{
In the ASOCC example,  {\dbname} selects \textit{no detection} for cold data, which corresponds to optimistic execution. It selects \textit{detect all} for hot data, forcing conflict detection for operations accessing highly contended records.
It selects \textit{detect critical} for warm data, enabling early validation to proactively  abort transactions that are likely to fail.}

\extended{
{{\dbname} implements several optimizations when maintaining this action lookup table, including pruning suboptimal functions from the search space and learning transaction backoff strategies.}

\mypar{Assign high priority after detect all}
Unlike 2PL, in {\dbname}, it is possible for an operation that passes the \textit{detect all} check to be ordered after another operation with a higher priority, leading to its eventual abort and retry even if both of them are using \textit{detect all} action.
Thus, we heuristically assign the highest priority to transactions that have passed the \textit{detect all} check, avoiding such unnecessary aborts.

\mypar{\textbf{Validate before expose}}
We notice that exposing dirty data must be done carefully, especially when combined with \textit{no detect} actions, as aborting a transaction that has exposed its data can result in a long chain of cascading aborts.
Thus, we follow existing work~\cite{DBLP:conf/osdi/WangDWCW0021/polyjuice, DBLP:conf/sigmod/WangMCYCL16/ic3, DBLP:conf/eurosys/MuAS19/DRP, DBLP:conf/sosp/XieSLAK015/Callas} to perform an early validation before exposing dirty data.

\mypar{\textbf{Learn transaction backoff}} We follow existing studies~\cite{DBLP:conf/sigmod/ShengTZP19/abort_prediction, DBLP:conf/osdi/WangDWCW0021/polyjuice} to learn a backoff for transaction abort and retry. This number is learned together with timeout, for each type of transaction.}

\extended{
\subsection{Function Update}
\label{sec:func_update}

The function is updated by optimizing a new function to replace the existing one when workload drifts are detected.
{\dbname} employs a simple, yet effective threshold-based method for workload drift detection.
Specifically, it monitors system performance and triggers the optimization process when detecting significant performance changes. The threshold for detection is 10\% (in throughput), which is determined empirically to ensure robust detection and to avoid over-triggering.  
Section~\ref{sec:evaluation.stored} shows that this detection method is effective.  We emphasize that {\dbname} only focuses on the design of the concurrency control algorithm, therefore the problem of workload drift detection is orthogonal. A detailed analysis of how our detection method compares to other algorithms, such as those discussed in ~\cite{DBLP:journals/pacmmod/WuI24/2sample}, is beyond the scope of the paper.

{\dbname} performs function updates without blocking the ongoing or incoming transactions. In particular, each transaction is assigned a specific function, namely the latest one, at the start of its execution.
During a function update, {\dbname} maintains both the latest function and the previously loaded functions still in use by ongoing transactions. While a new function is being loaded, the old function continues to handle transactions.
Once the loading is complete, the system performs an atomic pointer update to switch the latest function to the newly loaded one. 
This design avoids blocking ongoing or incoming transactions, and prevents switching functions in the middle of a transaction's execution.
}

\subsection{Transaction Execution}
\label{sec:details}
{After determining the CC actions, {\dbname} proceeds to execute the transaction operations. Similar to OCC, {\dbname} does not guarantee all conflicts are detected and resolved during execution. Instead, it performs a validation before committing the transaction, once all operations are executed. We elaborate on these steps below. 
}

\begin{algorithm}[t]
    \small
    \DontPrintSemicolon
	\caption{Execute an operation with learned actions
 }
	\label{alg:bdta.exection}
	\SetKwFunction{FMain}{Execute}
	\SetKwProg{Fn}{Function}{:}{}
	\Fn{\FMain{$T, op$}}{
        \textbf{a} $\gets$ GetCC($T, op$) \label{line:exe_get_action} // compute {\inputstate} and {\cacheF}({\inputstate}) \;
        {$op.priority \gets$ \textbf{a}$_{p}$\,,  $op.tx \gets T$\;}

        \If{\textbf{a}$_{d}$ = \textit{no detection} \label{line:exe_if_no_detect}}{
        $op.clean = True$, $S \gets \varnothing$ \label{line:exe_if_no_detect_set_clean} \; 
        }
        \ElseIf{\textup{\textbf{a}}$_{d}$ = \text{\textit{detect all}} \label{line:exe_if_detect_all}}{
        $S \gets $ all operations from conflict transactions \label{line:exe_add_all_conflict}
        }
        \ElseIf{\textup{\textbf{a}}$_{d}$ = \text{\textit{detect critical}} \label{line:exe_add_critical_conflict}}{
            \If{\textup{IsStoredProcedure()}}{
             $S \gets \varnothing$   \tcp*[f]{\footnotesize  {Conflicts arising from pipeline waits.}} \;
             \For{$T' \in T.dep$ \label{line:exe_start_add_safe_guards}}
                {bar $\gets$ $w_{T'.type}$ \\
                $S \gets S \cup \{ \text{the first bar-th accesses in } T' \}$\ \label{line:exe_end_schedule}
             }
            }
            \ElseIf{ \textup{EarlyValidation}$(T)$ = Abort \label{line:exe_early_validation_before} }{\Return{Abort, $\varnothing$}}
        }

        $S \gets S - \{op' \mid op' \in S \textup{ \textbf{and} } op'.priority < op.priority\}$ \label{line:exe_filter_by_rank} \;
        \If{Not all $op' \in S$ finish within timeout \textbf{a}$_{t}$  \label{line:exe_resolve_with_timeout}}{\Return{{Abort, $\varnothing$}} \label{line:exe_resolve_timeout_abort}}
        $result \gets $ SafeExecute($T, op$) \label{line:exe_operation} \;
        
        \If{\textup{IsStoredProcedure() \textbf{and} $expose=1$} \label{line:exe_if_expose}}{
             \If{ \textup{EarlyValidation}$(T)$ = Abort \label{line:exe_early_validation2}}{\Return{{Abort, $\varnothing$}}}
             \Else{
                \extended{
                \If{$op$ is not the last operation in $T$ \label{line:alg.exec.defer_expose_start}}{
                    $\mathbf{a}' \gets $ the actions for the next operation in $T$\;
                    Perform the pipeline wait actions in $\mathbf{a}'$ \;
                    }\Else{
                        Wait for all $T' \in T.dep$ to finish \label{line:alg.exec.defer_expose_end}\;
                    }
                }
                \lFor{$op' \in T.ws$}{Lock tuple($op'.key$)}
                \For{$op' \in $ \textup{dirty reads on tuple($op.key$)}}{
                    $T.dep \gets T.dep \cup \{op'.tx\}$ \label{line:exe_add_wr_dependency}
                }
                \For{$op' \in T.ws$}{
                    Append $op'$ to the version chain as dirty data \label{line:exe_add_dirty} \;
                    Unlock tuple($op'.key$) \label{line:exe_end_of_expose}
                }
             }
        }
		\Return{{Succeed, result}}\;
	}

 \SetKwFunction{FMain}{SafeExecute}
	\SetKwProg{Fn}{Function}{:}{}
	\Fn{\FMain{$T, op$}}{
        \If{$op.type = $ read}{
            \If{$op.clean$}{
                $op' \gets$ latest committed data on tuple($op.key$) \label{line:exe_clean_read} \;
            }\Else{
                $op' \gets$ latest dirty data on tuple($op.key$) \label{line:exe_dirty_read} \;
                $T.dep \gets T.dep \cup \{op'.tx\}$ \label{line:exe_add_rw_dep}
            }
            $T.rs \gets T.rs \cup \{op'\}$  \label{line:exe_add_read_set} \;
            $T.validate \gets T.validate \cup \{op'\}$  \label{line:exe_add_validation_set}  \;
            \Return{$op'$}
        }\Else{
            {$T.ws \gets T.ws \cup \{op\}$ \label{line:exe_add_write_set} \;}
            \Return{$\varnothing$}
        }
    }
\end{algorithm}

\subsubsection{Execution}
\label{sec:details.execution}
A transaction in {\dbname} has the following fields:
\begin{itemize}
    \item $T.rw$ and $T.ws$: the local read and write sets.
    \item $T.dep$: the set of transactions that $T$ depends on.
    \item $T.type$: the transaction type of $T$.
    \item $T.validate$: operations of $T$ that have been executed but not yet validated.
\end{itemize}
{For each operation $op$, {\dbname} maintains the following information:
\begin{itemize}
    \item $op.tx$: the transaction that contains $op$.
    \item $op.key$: the unique key of the data object accessed by $op$.
    \item $op.type$: the operation type (read/write).
    \item $op.clean$: whether $op$ reads the latest committed (clean) data.
    \item $op.priority$: the priority of $op$ when resolving its conflicts with other operations.
\end{itemize}
}

{Algorithm~\ref{alg:bdta.exection} describes how {\dbname} executes an operation.
It performs early validation to proactively abort transactions that will fail later (line ~\ref{line:exe_early_validation_before}).
It also performs early validation before exposing a write (line~\ref{line:exe_early_validation2}), where the failure of a transaction $T$ could lead to the cascading abort of all transactions depending on it. 
\extended{By performing early validation (Algorithm~\ref{alg:validation}), it reduces the impact of cascading aborts.}
The algorithm does not perform early validation for ``detect all'' because doing so incurs additional overhead and can restrict the search space. For example, it would prevent {\dbname} from replicating 2PL, which does not incur the overhead of early validation checks.}

\revisionhint{F.Q1}{{\dbname} does not outperform 2PL when all the actions are detect-all, nor does it introduce a new lock-based protocol.
However, we note that despite the extra validation cost compared to classical algorithms such as 2PL, the performance gained from the learned concurrency control actions outweighs this cost and leads to higher throughput, as will be illustrated in Section~\ref{sec:evaluation}.} 
\revisionhint{R2.O1}{Furthermore, when all actions in the agent function are \textit{detect all}, {\dbname} can skip validation mechanisms, including the tracking of $T.rs, T.ws, T.validate$ and the transaction validation at commit time, to further reduce the overhead over 2PL.}

The algorithm distinguishes between stored procedure and interactive transaction modes. In the former, all the data accesses are known prior to transaction execution. In the
latter, they are not known before transaction execution.
For interactive transactions, the algorithm disallows actions that could violate correctness.  Such actions include dirty reads and partial retry, which are disallowed because the behavior of the conflicting transaction is unknown, and because the user has already seen the outputs of some operations.  

Given an operator $op$,
the algorithm first invokes $GetCC(T, op)$ to get the action $\action$ (line~\ref{line:exe_get_action}).
Next, it executes conflict detection and stores the conflicts in $S$.
This step is skipped if ${\action}_d$ is \textit{no detection}, and the algorithm enforces $op$ to read clean data (lines~\ref{line:exe_if_no_detect}--\ref{line:exe_if_no_detect_set_clean}). 
If ${\action}_d$ is \textit{detect critical}, it performs early validation to check if previously read data have changed
(line~\ref{line:exe_early_validation_before}) and aborts $T$ if the validation fails.
The conflicts with lower priorities than that of $op$ are removed from $S$. The remaining are resolved with the given
timeout ${\action}_t$ (lines~\ref{line:exe_filter_by_rank}--\ref{line:exe_resolve_with_timeout}). If not all the conflicts in
$S$ are resolved within ${\action}_t$, $T$ is aborted (line~\ref{line:exe_resolve_timeout_abort}). Otherwise, the algorithm proceeds to read the latest committed data (line~\ref{line:exe_clean_read}), or to write to its write set $T.ws$ (line~\ref{line:exe_add_write_set}).
For stored procedure transactions, the algorithm allows dirty reads.
\revisionhint{R2.O2}{In particular, it uses $w_1, \dots,
w_n$ to detect critical conflicts before dirty reads, uses $expose$ to decide whether to make uncommitted write visible
to other transactions, and records dependencies caused by dirty reads in $T.dep$ (lines~\ref{line:exe_add_wr_dependency} and~\ref{line:exe_add_rw_dep}).}
Each data object has a version chain containing dirty data.
The execution when ${\action}_d$ is \textit{no detection} or \textit{detect all} is the same as that of interactive transactions.
When ${\action}_d$ is \textit{detect critical}, it detects critical conflicts between $T$ and its dependent transactions in $T.dep$ (lines
~\ref{line:exe_start_add_safe_guards}--\ref{line:exe_end_schedule}).
In particular, for every $T' \in T.dep$, it adds the first $w_{T'.type}$ accesses of $T'$ to $S$.
Once all the conflicts in $S$ are resolved, it lets $op$ read the data at the tail of the corresponding version chain, and updates $T.dep$ to track the read-write dependency (lines~\ref{line:exe_dirty_read}--\ref{line:exe_add_rw_dep}).
After executing $op$ (line~\ref{line:exe_operation}), if $expose = 1$, it performs early validation and aborts $T$ if the validation fails.
If the validation succeeds, it exposes the uncommitted data in $T.ws$ by appending them to the corresponding version chains. 

\revisionhint{R1.O1 R1.O2}{
We use the ASOCC example to illustrate how a transaction is executed with \dbname. We assume the transaction is in the interactive transaction mode. 
Before executing an operation, {\dbname} first computes the input state $\inputstate$  and retrieves the corresponding actions by evaluating $\cacheF(\inputstate)$ (line~\ref{line:exe_get_action}).
If the selected action is \textit{no detection} (i.e., the accessed data is cold), {\dbname} executes it optimistically without blocking
(line~\ref{line:exe_if_no_detect}).
If the action is \textit{detect all} (i.e., the accessed data is hot), {\dbname} performs conflict detection before execution
(line~\ref{line:exe_if_detect_all}), and upon detecting a conflict, it blocks or abort the transaction according to the selected timeout (line~\ref{line:exe_resolve_with_timeout}).
If the action is \textit{detect critical} (i.e., the accessed data is warm), {\dbname} performs early validation (line~\ref{line:exe_add_critical_conflict}), and if validation fails, it aborts the transaction immediately.
}

\maintext{

\subsubsection{Validation and commit}
{\dbname} performs a four-step validation at commit time, similar to OCC.  First, it waits for all transactions in $T.dep$ to complete, and aborts if any dependent transaction aborts. Second, it locks the private write set to avoid concurrent reads of the data in the set. 
Third, it verifies that data in the read set matches the latest committed versions, and aborts if the verification fails.
Finally, it removes dirty data from $T$ on the corresponding version chains and unlocks the write set. The latest committed data are updated
with those in the write set.
{We present detailed validation and commit algorithms in the extended version~\cite{NeurCC-extended}.}

\subsubsection{Correctness}
{
Similar to OCC, {\dbname} ensures correctness (i.e., serializability) by performing validation before committing the transaction. A detailed proof of correctness is provided in the extended version~\cite{NeurCC-extended}.
}
}

\extended{
\begin{algorithm}[t]
    \small
    \DontPrintSemicolon
	\caption{Validation and commit}
	\label{alg:validation}
	\SetKwFunction{FMain}{EarlyValidation}
	\SetKwProg{Fn}{Function}{:}{}
	\Fn{\FMain{$T$}}{
         \For{$op \in T.validate$ \label{line:exe_start_early_validation}}{
        $op' \gets $ latest data version on tuple($op.key$). \;
        \If{$op.tx \neq op'.tx$}{\Return{Abort}}
         }
         $T.validate = \varnothing$ \;
         \Return{Succeed}
    }
	\SetKwFunction{FMain}{Commit}
	\SetKwProg{Fn}{Function}{:}{}
	\Fn{\FMain{$T$}}{
         \For{$T' \in T.dep$ \label{line:valid_wait_dep}}{
        Wait $T'$ to finish. \;
        \If{$T$ read dirty from $T'$ \textbf{\textup{and}} $T'$ gets aborted}{\Return{Abort} \label{line:valid_wait_dep_end}}
         }
        \lFor{$op \in T.ws$ \label{line:valid_lock_write_set}}{
        Lock tuple($op.key$).
         }
         *\tcp{\footnotesize Serialization point. \label{line:valid_serialization_point}}
        \For{$op \in T.rs$ \label{line:valid_check_read_set}}{
         $op' \gets $ latest committed data version on tuple($op.key$)  \;
         \lIf{$op.tx \neq op'.tx$ \textbf{\textup{or}} $op$ is locked by other transactions}{
         {\Return{Abort} }
         \label{line:valid_check_read_set_end}}
         }
        \For{$op \in T.ws$ \label{line:valid_unlock_write_set}}{
         Clean up $T$'s dirty data on tuple($op.key$)\;
         Append $op$ to tuple($op.key$)'s version chain as clean data \label{line:valid_add_write}\;
         Unlock tuple($op.key$). \label{line:valid_unlock_write_set_end}
         }
         \Return{Succeed}
	}
\end{algorithm}

\subsubsection{Validation and commit}
Algorithm~\ref{alg:validation} shows how the algorithm performs an OCC-like four-step validation at commit time.  First,
it waits for all dependent transactions in $T.dep$ to complete, and aborts if any dependent transaction aborts 
(lines~\ref{line:valid_wait_dep}--\ref{line:valid_wait_dep_end}). Second, it locks the private write set to avoid
concurrent read of the data in the set (line~\ref{line:valid_lock_write_set}). 
Third, it verifies that data in the read
set are the same as the latest committed versions
(lines~\ref{line:valid_check_read_set}--\ref{line:valid_check_read_set_end}). It aborts if the verification fails.
Fourth, it removes dirty data from $T$ on the corresponding version chains and unlocks the write set
(lines~\ref{line:valid_unlock_write_set}--\ref{line:valid_unlock_write_set_end}). The latest committed data is updated to those in the write set. 

\subsection{Proof of Correctness}

To prove that {\dbname} ensures serializability, we show that any transaction schedule allowed by {\dbname} can be transformed into a serial schedule in which the committed transactions are executed one by one in isolation. In particular, we show that the transaction execution results are the same when running with the {\dbname} schedule as when running with the serial schedule.
\\

\mypar{Definitions and Setup} 
Let:
\begin{itemize}
    \item Each transaction \( T \) is assigned a unique timestamp $t(T)$ at its serialization point (Algorithm~\ref{alg:validation}, line~\ref{line:valid_serialization_point}).
    \item Let \( T_1, T_2, \dots, T_n \) be the set of committed transactions, indexed such that \( t(T_1) < t(T_2) < \cdots < t(T_n) \), where \( t(T_i) \) denotes the serialization point of transaction \( T_i \).
    \item A \emph{schedule} \( S \) is an interleaved sequence of read and write operations from  \( T_1, T_2, \dots, T_n \).
    \item Two schedules of committed transactions are \emph{equivalent} if, for every committed transaction, all of its read operations return the same values in both schedules, and the two schedules result in the same final database state.
    \item A schedule is \emph{serial} if all operations of each transaction execute contiguously, without interleaving with other transactions.
    \item A schedule is \emph{serializable} if it is equivalent to some serial schedule.
    \item Let schedule $S_{k}$ be a subsequence of $S$ containing all operations from transactions \(T_1, T_2, \dots, T_k\) (\(S_n = S\)). 
    \item Let schedule $S'_{k}$ be a serial schedule $\langle T_1, T_2, \dots, T_k \rangle$ in which $T_1, T_2, \dots T_k$ are executed sequentially.
\end{itemize}

We aim to prove that any schedule \(S\) produced by {\dbname} is serializable. That is, \(S_{n}\) is equivalent to a serial schedule \(S'_{n}\) in which committed transactions \(T_1, T_2, \dots, T_n\) are executed sequentially. We prove by induction on \(k\) (from \(1\) to \(n\)) that \(S_k\) is equivalent to \(S'_k\).

\mypar{Base Case (k = 1)} For a single committed transaction \( T_1 \), the schedule \( S_1 \) trivially corresponds to the serial schedule \( \langle T_1 \rangle \), as there is no transaction operation interleaving.

\mypar{Inductive Step}  Assume that for \( T_1, T_2, \dots, T_k \), the schedule \( S_k \) is equivalent to the serial schedule \(S'_{k}\). Now consider an additional committed transaction \( T_{k+1} \), with \( t(T_k) < t(T_{k+1}) \). We show that inserting operations from $T_{k+1}$ to $S_k$ to form $S_{k+1}$ still produces the same return values and final state as the schedule $S'_{k+1}$.

Let \( r(x) \) denote a read operation on data object \( x \) from transaction \( T_{k+1} \), \(w(x)\) denote a write operation on data object \( x \) from transaction \( T_{k+1} \). We now prove that $S_{k+1}$ and $S'_{k+1}$ produce the same \emph{return values} and \emph{final database state}.

\begin{itemize}
    \item \textbf{\(T_{1}, \dots, T_{k}\) read the same return values in \(S_{k+1}\) and \(S'_{k+1}\):} For any transaction \(T_i \in \{T_{1}, \dots, T_{k}\}\), all of its read operations occur before its serialization point \(t(T_i)\). Since any write \(w(x) \in T_{k+1}\) becomes visible only after \(t(T_{k+1})\), and \(t(T_i) < t(T_{k+1})\), none of the reads in \(T_i\) can observe writes from \(T_{k+1}\). Furthermore, since \(T_{1}, \dots, T_{k}\) read the same return values in \(S_k\) and \(S'_k\), and the state visible to them remains unchanged considering updates from $T_{k+1}$, they continue to read the same values in $S_{k+1}$ and $S'_{k+1}$.
    
    \item \textbf{\(T_{k+1}\) reads the same return values in \(S_{k+1}\) and \(S'_{k+1}\):} Since \(t(T_{k+1}) > t(T_i)\) for all \(i \in \{1, 2, \dots, k\}\), any read operation \(r(x) \in T_{k+1}\) occurs after the updates to \(x\) from $S_{k}$. Otherwise, \(T_{k+1}\) would detect a conflicting write to $x$ from a transaction $T_i$ with $t(T_i)<t(T_{k+1})$, and abort due to a failed validation of its read set (Algorithm~\ref{alg:validation}, line~\ref{line:valid_check_read_set_end}). In summary, each \(r(x) \in T_{k+1}\) observes the final state of \(x\) resulting from all updates in \(S_k\). Since the final database states produced by \(S_k\) and \(S'_k\) are identical, and \(T_{k+1}\) reads from this final state in both \(S_{k+1}\) and \(S'_{k+1}\), \(T_{k+1}\) reads the same return values in \(S_{k+1}\) and \(S'_{k+1}\).

    \item \textbf{\(S_{k+1}\) and \(S'_{k+1}\) produce the same final database state:} Since \(t(T_{k+1}) > t(T_i)\) for all \(i \in \{1, 2, \dots, k\}\), any update operation \(w(x) \in T_{k+1}\) occurs after the updates to \(x\) from $S_{k}$. Otherwise, \(w(x)\) would have been blocked during write lock acquisition (Algorithm~\ref{alg:validation}, line~\ref{line:valid_add_write}). This guarantees that no update in \(S_k\) overwrites or interferes with the updates made by \(T_{k+1}\), ensuring that \(S_{k+1}\) produces the same final state as \(S'_{k+1}\) for all \(x\) modified by \(T_{k+1}\). For data objects not modified by \(T_{k+1}\), the final state is already identical in \(S_k\) and \(S'_k\), so the resulting state remains the same in both \(S_{k+1}\) and \(S'_{k+1}\).
\end{itemize}

By induction, for all $k \in [1, n]$, we have $S_k$ equivalent to $S'_k$. Hence, the full schedule $S = S_n$ is equivalent to the serial schedule $S'_n$. Therefore, {\dbname} ensures serializability.

}

\section{Learning Function and Feature Selector}
\label{sec:fine-tuning}

{
In this section, we describe how {\dbname} learns the function {\cacheF}.
We observe that different actions are amenable to different optimization techniques.}
More specifically, model-based optimizations are
suitable for some actions, whereas search-based techniques are suitable for others.  Furthermore, different actions have
different impacts on the overall throughput, and thus the order in which they are optimized can affect convergence speed.
Based on these observations, we decompose {\cacheF} into three sub-functions: {\cacheFT}, {\cacheFD}, and {\cacheFP},
and optimize them separately. Their outputs are combined to obtain {\action} $ = \langle {\action}_d, {\action}_t,
{\action}_p, ({w_1}, \dots, {w_n}, {expose}) \rangle$.  {\cacheFT} controls the timeout values, corresponding to ${\action}_t$.
{\cacheFD} controls the conflict detection and
priority, corresponding to (${\action}_d, {\action}_p$). {\cacheFP} controls pipeline-wait actions, corresponding to $({w_1},
\dots, {w_n}, {expose})$.

Learning the optimal function requires evaluating the performance  of {\cacheF}, i.e., Score({\cacheF}). Existing works~\cite{DBLP:conf/usenix/TangE18/CormCC, DBLP:conf/osdi/WangDWCW0021/polyjuice, DBLP:conf/cidr/TangJE17/ACC} load the function into the running database and use the system throughput as Score({\cacheF}). 
This approach is time-consuming, because loading and running the system takes seconds\footnote{This process takes 3.68 seconds on average under our default experimental setting, as detailed in Section~\ref{sec:evaluation.setup}: TPCC workload, 16 threads, one warehouse.} to complete, and they need to be repeated at every optimization step, adding significant overheads to the learning process~\cite{DBLP:conf/osdi/WangDWCW0021/polyjuice}.
In addition, Score({\cacheF}) is non-smooth due to its discrete input {\cacheFP}, which makes it difficult to optimize {\cacheFP} with model-based optimization methods. {\dbname} addresses the aforementioned limitations of the existing approaches for evaluating Score({\cacheF}) in two ways. First, it employs an online-trained surrogate model to minimize the number of performance evaluations (Section~\ref{sec:bayesian_opt}). Second, it uses a graph reduction search algorithm to effectively optimize {\cacheFP} (Section~\ref{sec:graph_reduction}).

\subsection{Learning Function Via Surrogate Model}
\label{sec:bayesian_opt}

As Score({\cacheF}) is costly to evaluate, {\dbname} treats it as a black-box~\cite{regis2013combining/DYCORS, DBLP:conf/kdd/GolovinSMKKS17/Vizier, DBLP:books/daglib/0022247} and adopts a Bayesian optimization approach. Bayesian optimization can efficiently identify high-performing functions with a small number of evaluations, thereby reducing the cost.
In particular, {\dbname} employs a probabilistic surrogate model, i.e., the Gaussian process model, which is trained online to approximate Score({\cacheF}).
This model is differentiable for efficient optimization and is well-suited as a surrogate model because it predicts both the mean and variance of the performance, which balances exploration of new areas in the function space with exploitation of known high-performing areas.

The surrogate model guides the optimization process by identifying the most promising 
function, i.e., the one expected to yield the highest performance  improvement, 
for subsequent deployment and evaluation.
Specifically, during an optimization step, the surrogate model is first {fitted to the historical data collected  since the last workload drift}, which optimizes its mean vector and covariance matrix to maximize the data likelihood.
Next, for an unevaluated function $\cacheF$, the fitted model predicts the mean $\mu(\text{Score}(\cacheF))$ and variance $\sigma(\text{Score}(\cacheF))$ of its performance.
After that, the Upper Confidence Bound (UCB) acquisition function computes a score for $\cacheF$ as $\text{UCB}(\cacheF) = \mu(\text{Score}(\cacheF)) + \lambda \sigma(\text{Score}(\cacheF))$, where $\lambda$ is a trade-off parameter. It is set to $2.576$ in our experiments, corresponding to the 99th percentile of a standard normal distribution to achieve an effective balance between exploration and exploitation.
Note that the UCB score is differentiable with respect to the parameters of $\cacheF$, and therefore {\dbname} can explicitly optimize these parameters to maximize the UCB score using the L-BFGS-B~\cite{DBLP:journals/mp/LiuN89} algorithm, which produces $\cacheF_{next}$, the function with the highest expected performance.
Once $\cacheF_{next}$ is identified, it is evaluated in the database system, and its actual performance score is recorded.
Finally, the new data point of $\cacheF_{next}$ and its score are added to the historical data, and the surrogate model is updated.
This process is repeated, progressively refining predictions of Score({\cacheF}) and identifying better functions over time.

\subsection{Pipeline-Wait Action Optimization via Learned Conflict Graph}
\label{sec:graph_reduction}
The Bayesian optimization technique in Section~\ref{sec:bayesian_opt} can effectively optimize {\cacheFT} and {\cacheFD}, but optimizing {\cacheFP} is more challenging, especially for complex workloads like TPC-C that involve many potential dependency cycles. 
{We observe empirically that during the optimization of {Score}({\cacheFT}, {\cacheFD}, \cacheFP), the function exhibits poor locality with respect to {\cacheFP}} around the optimal solutions.
Even minor changes to {\cacheFP} can significantly reduce performance by introducing new dependency cycles that lead to cascading aborts or deadlocks~\cite{DBLP:conf/sigmod/WangMCYCL16/ic3}.
As a result, the
surrogate model trained via Bayesian optimization fails to approximate Score({\cacheFT}, {\cacheFD}, \cacheFP)
accurately, because the optimization inherently assumes smoothness and locality in the approximated function.

\extended{

\begin{algorithm}[t]
    \small
    \DontPrintSemicolon
    \caption{{Graph reduction search algorithm}}
    \label{alg:conflict_graph_training}

    \SetKwFunction{FTrain}{GraphReduction}
    \SetKwProg{Fn}{Function}{:}{}
    \Fn{\FTrain{$branchFactor, mutateRate, K$}}{
        $\modiV_{\textup{false}} \gets $ a default function that returns \textbf{false} for all inputs \;
        {${\cacheF}_{best} \gets $ the current best function.} \;
        $Pop \gets \{ (\modiV_{\textup{false}}, \modiV_{\textup{false}}, initialScore)\}$ \;
        \While{$Pop \neq \varnothing$ \label{line:alg.graph_reduct.stop_condition_2}}{
            \For{$(\modiV_{e}, \modiV_{o}, score) \in Pop$}{
                \For{$i \gets 1$ \KwTo branchFactor \label{line:alg.graph_reduct.branch_start}}{
                    $(\modiV_{e}', \modiV_{o}', score') \gets \textup{Mutate(}\modiV_{e}, \modiV_{o}, mutateRate$\textup{)} \;
                    $nTry \gets 1$ \;
                    \While{\textup{(}$\modiV_{e}', \modiV_{o}'$\textup{) has been searched before} \textbf{and} $nTry \le MaxTry$}{
                        $(\modiV_{e}', \modiV_{o}', score') \gets \textup{Mutate}(\modiV_{e}, \modiV_{o}, mutateRate)$ \;
                        $nTry \gets nTry + 1$
                    }
                    \If{$nTry > MaxTry$}{ 
                        $Pop \gets Pop - \{(\modiV_{e}, \modiV_{o}, score)\}$\;
                        \textbf{break}
                    }\Else{
                        $Pop \gets Pop \cup \{(\modiV_{e}', \modiV_{o}', score')\}$\;
                    }
                }
                \label{line:alg.graph_reduct.branch_end}
            }
            $Pop \gets$ GetTopK($Pop$) \;
            \If{$Pop$ did not change in $MaxStep$ steps \label{line:alg.graph_reduct.stop_condition_1}}{
                \textbf{break}
            }
        }
        \Return ${\cacheF}_{best}$\;
    }
    
    \SetKwFunction{FMutate}{Mutate}
    \SetKwProg{Fn}{Function}{:}{}
    \Fn{\FMutate{$\modiV_{e}, \modiV_{o}, prob$}}{
        \tcp{\footnotesize {Mutate by merging nodes.}}
        \For{$((T, op) \to expose) \in \modiV_e$ \label{line:alg.graph_reduct.mutate_start}}{
            \text{Mutate} $expose$ \text{to} \textbf{true} {with $prob$ probability}\;
        }
        \tcp{\footnotesize {Mutate by cutting edges.}}
        \For{$((T, op) \to optimistic) \in \modiV_o$}{
            \text{Mutate} $optimistic$ \text{to} \textbf{true} {with $prob$ probability}\;
        }
        ${\cacheF}_{new} \gets \text{UpdateWaitActions}(\modiV_{e}, \modiV_{o}, {\cacheF}_{best})$ \;
        \If{$\textup{Score}({\cacheF}_{new}) > $ \textup{Score(}${\cacheF}_{best}$\textup{)} \label{line:alg.graph_reduct.evaluate}}{
            ${\cacheF}_{best} \gets {\cacheF}_{new}$ \;
        }
        \Return $\modiV_e, \modiV_o, \text{Score}({\cacheF}_{new})$ \label{line:alg.graph_reduct.mutate_end} \;
    }
    
\end{algorithm}

}

To address the challenge of global optimization of {\cacheFP}, {\dbname} learns an optimal conflict graph instead of learning these actions directly.
\revisionhint{R2.O3}{The optimal conflict graph is the learned conflict graph from which
the pipeline-wait actions that maximize the performance score are derived.}
This reformulation simplifies the
learning process and improves optimization performance.
A conflict graph is a 
graph generated from the workload, where each type of transaction access (e.g., a line of SQL statement) is represented as a node. If two operations have potential conflicts (e.g., read-write conflicts), an edge is added between the corresponding nodes. {\dbname} uses SC-graph, a conflict graph widely adopted in existing CC algorithms~\cite{DBLP:conf/osdi/WangDWCW0021/polyjuice, DBLP:conf/sigmod/WangMCYCL16/ic3, DBLP:conf/osdi/MuCZLL14/Rococo}.
These algorithms construct a static conflict graph and compute pipeline-wait actions by analyzing this graph before workload execution.
However, they rely
on static conflict graphs, which do not account for dynamic factors such as access skewness.
We contend that the conflict graph should be dynamic, with conflict edges added based on certain probabilities.
For example, in the TPC-C workload, when there are many warehouses and few concurrent users, the \textit{payment} transaction's update WAREHOUSE table access is unlikely to conflict with the \textit{new order} transaction’s read WAREHOUSE table
access, and therefore this edge can be deleted. However, when there are few warehouses and many concurrent users,
these accesses are more likely to conflict, therefore adding the conflict edge is beneficial.

\maintext{
\revisionhint{R2.O3}{{\dbname} employs a graph reduction search algorithm (Algorithm 3 in the extended version~\cite{NeurCC-extended}) to efficiently learn the optimal conflict graph.
{Given an initial static conflict graph}, the algorithm keeps track of the top-$K$ best-performing conflict graphs found during the search.
Each graph is evaluated by converting it into pipeline-wait actions and evaluating the corresponding function.
Specifically, the algorithm first performs conflict graph analysis using an IC3-like algorithm~ \cite{DBLP:conf/sigmod/WangMCYCL16/ic3} 
to find potential dependency cycles and to compute the pipeline-wait actions that resolve them (see Algorithm 4 in the extended version~\cite{NeurCC-extended}).
It next updates the current best agent function with the computed pipeline-wait actions (Algorithm 5 in the extended version~\cite{NeurCC-extended}). Finally, the algorithm evaluates the updated agent function and uses its performance as the score of the conflict graph (Algorithm 3 in the extended version~\cite{NeurCC-extended}).
}}

\maintext{
\begin{figure}[htbp]
  \centering
  \includegraphics[width=0.8\linewidth]{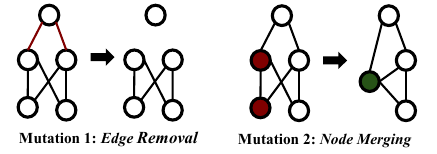}
  \vspace{-5mm}
  \caption{Two types of graph reduction mutations.}
  \vspace{-1mm}
  \Description{Two types of graph reduction mutations.}
  \label{fig:mutation}
\end{figure}}

\extended{

\begin{algorithm}[htbp]
    \small
    \DontPrintSemicolon
    \caption{{Fetching pipeline-wait actions}}
    \label{alg:wait_actions}
    
    \SetKwFunction{FMain}{GetWaitActions}
    \SetKwProg{Fn}{Function}{:}{}
    \Fn{\FMain{$op, T, \modiV_e, \modiV_o, LastExposed$}}{
        $G_{\text{cut}} \gets \text{CutEdgesAroundOptimisticOperations}(G_{\text{full}}, \modiV_o)$ \label{line:alg.get_wait.calculate_wait_actions_start} \;
        $n \gets$ number of operations in $T$\;
        $lastExposePoint \gets n+1$\;
        $wait_{op, T.\text{type}} \gets 0$\;
        \tcp{\footnotesize {Get wait actions.}}
        \For{$i \gets n$ \KwTo 1}{
            $op' \gets$ the $i_{\text{th}}$ operation of $T$\;
            \If{$\modiV_e(T,op') = \textbf{\textup{false}}$}{$lastExposePoint \gets i$\;}
            \If{$\langle op, op' \rangle \in G_{\text{cut}}$ \textbf{and} $op'$ is read}{
                $wait_{op, T.\text{type}} \gets i$\;
                \textbf{break}
            }
            \If{$\langle op, op' \rangle \in G_{\text{cut}}$ \textbf{and} $op'$ is write}{
                $wait_{op, T.\text{type}} \gets lastExposePoint$\;
                \textbf{break}         \label{line:alg.get_wait.calculate_wait_actions_end}
            }
        }
        \tcp{\footnotesize {Defer the wait actions of write operations until they are exposed to concurrent transactions.}}
        \If{$op$ is read \label{line:alg.get_wait.defer_expose_wait_start}}{
            $res \gets wait_{op, T.\text{type}}$\;
        }\Else{
            $res \gets 0$\;
        }
        \If{$LastExposed \neq \varnothing$} {
            \For{$op' \in LastExposed$}{
                \If{$wait_{op', T.\text{type}} > res$ \textbf{and} $op'$ is write}{
                    $res \gets wait_{op', T.\text{type}}$ \label{line:alg.get_wait.defer_expose_wait_end}\;
                }
            }
        }
        \Return $res$\;
    }
\end{algorithm}

{\dbname} employs a graph reduction search algorithm inspired by genetic search to quickly learn the optimal conflict
graph.
{The algorithm starts with a large static conflict graph $G_{full}$, where each node represents a transaction operation $op$ from a transaction $T$, and an edge connecting two operations accessing the same table and least one operation is a write.
The modifications of the conflict graph $G_{full}$ are defined by two functions, \( \modiV_e \) and \( \modiV_o \):
\begin{itemize}
    \item \( \modiV_e(T, op) = \) true if the node corresponding to operation \( op \) in \( T \) is merged with the node corresponding to the operation next to \( op \) in \( T \).
    \item \( \modiV_o(T, op) = \) true if edges involving the node corresponding to \( op \) in \( T \) are removed.
\end{itemize}

\begin{algorithm}[htbp]
    \small
    \DontPrintSemicolon
    \caption{{Updating pipeline-wait actions}}
    \label{alg:get_updated_agent}

    \SetKwFunction{FMain}{UpdateWaitActions}
    \SetKwProg{Fn}{Function}{:}{}
    \Fn{\FMain{$\modiV_e, \modiV_o, {\cacheF}_{best}$}}{
        \tcp{\footnotesize {Get pipeline wait actions for all transaction operations.}}
        \For{$T \in $ all types of transactions}{
            $n \gets$ number of operations in $T$\;
            \For{$T' \in $ all types of transactions}{
                $unexposed \gets \varnothing$ \;
                $toExpose \gets \varnothing$ \;
                \For{$i \gets 1$ \KwTo $n$}{
                    $op \gets$ the $i_{\text{th}}$ operation of $T$\;
                    $WaitAct_{T, T', i} \gets $ GetWaitActions($op, T', \modiV_e, \modiV_o, toExpose$) \;
                    $unexposed = unexposed \cup \{ op \}$\;
                    $toExpose = \varnothing$\;
                    \If{$\modiV_e(T, op) = $ \textup{\textbf{false}}}{
                        $toExpose = unexposed$\;
                        $unexposed \gets \varnothing$ \;
                    }
                }
            }
        }
        $\mathbf{S} \gets $ GetInputSet(${\cacheF}_{best}$)\;
        $\cacheF_{new} \gets {\cacheF}_{best}$\;
        \tcp{\footnotesize {Update the pipeline wait actions in $\cacheF_{new}$.}}
        \For{$s \in \mathbf{S}$}{
            $isOptimistic \gets $ \textup{\textbf{true}}\;
            \For{$T' \in $ all types of transactions}{
                $\cacheF_{new}(s).w_{T'.type} \gets 0$ \;
                \For{$T$ whose $i_{th}$ operation can produce $s$ state}{
                    $op \gets$ the $i_{\text{th}}$ operation of $T$\;
                    \If{$WaitAct_{T, T', i} > \cacheF_{new}(s).w_{T'.type}$ \label{line:alg.update_wait.take_max_wait_start}}{
                        $\cacheF_{new}(s).w_{T'.type} \gets WaitAct_{T, T', i}$ \label{line:alg.update_wait.take_max_wait_end} \;
                    }
                    \If{$\modiV_o(T,op) = $ \textup{\textbf{false}}}{
                        $isOptimistic \gets $ \textup{\textbf{false}} \;
                    }
                }
            }
            \If{$isOptimistic = $ \textup{\textbf{true}} \label{line:alg.update_wait.isolated_node_start} }{
                        $\cacheF_{new}(s).\mathbf{a}_d \gets $ \textit{no detect} \label{line:alg.update_wait.isolated_node_end}
                    }
        }
        \Return ${\cacheF}_{new}$\;
    }
\end{algorithm}

\begin{figure}[htbp]
  \centering
  \includegraphics[width=0.82\linewidth]{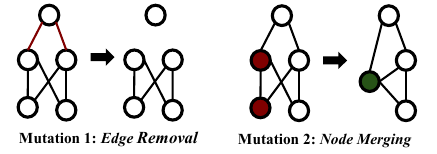}
  \vspace{-4mm}
  \caption{Two types of graph reduction mutations.}
  \Description{Two types of graph reduction mutations.}
  \label{fig:mutation}
\end{figure}

As outlined in Algorithm~\ref{alg:conflict_graph_training}, the graph search algorithm starts with no conflict graph modifications, \( \modiV_e = \modiV_o = \modiV_{\text{false}} \), then searches for better conflict graphs through mutations (lines \ref{line:alg.graph_reduct.branch_start} -- \ref{line:alg.graph_reduct.branch_end} and lines \ref{line:alg.graph_reduct.mutate_start} -- \ref{line:alg.graph_reduct.mutate_end}).
The algorithm maintains a population of the top $K$ best conflict graphs found during the search.
The conflict graphs are evaluated by converting them into pipeline-wait actions (Algorithm~\ref{alg:wait_actions}), generating new functions with these pipeline-wait actions (Algorithm~\ref{alg:get_updated_agent}), and evaluating the corresponding functions.}
Specifically, given a conflict graph, we identify all potential dependency cycles and compute the required pipeline-wait actions to resolve these cycles (Algorithm~\ref{alg:wait_actions} lines \ref{line:alg.get_wait.calculate_wait_actions_start} -- \ref{line:alg.get_wait.calculate_wait_actions_end}). After this, we update the pipeline-wait actions in the current best function to our calculated pipeline-wait actions and evaluate its performance score as the score for the conflict graph (Algorithm~\ref{alg:conflict_graph_training} line \ref{line:alg.graph_reduct.evaluate}).
{As a function input may map to multiple nodes in the conflict graph, we select the longest wait action to prevent potential dependency cycles (Algorithm~\ref{alg:get_updated_agent} lines \ref{line:alg.update_wait.take_max_wait_start} -- \ref{line:alg.update_wait.take_max_wait_end}).
We adopt the same optimizations as IC3~\cite{DBLP:conf/sigmod/WangMCYCL16/ic3} to enforce the pipeline-wait actions of the next operation before exposing dirty writes (Algorithm~\ref{alg:bdta.exection} lines \ref{line:alg.exec.defer_expose_start} -- \ref{line:alg.exec.defer_expose_end}), and to defer the pipeline waits of unexposed write operations until they are exposed (Algorithm~\ref{alg:wait_actions} lines \ref{line:alg.get_wait.defer_expose_wait_start} -- \ref{line:alg.get_wait.defer_expose_wait_end}).}
}

During optimization, the algorithm iteratively improves the conflict graphs by mutating the current graphs to better capture actual conflicts.
Each mutation simplifies the graph by edge removal or by merging consecutive nodes  (as illustrated in Figure~\ref{fig:mutation}\extended{ and Algorithm~\ref{alg:conflict_graph_training}}). The mutation aims to eliminate unnecessary conflicts that will be resolved during the
pipeline-wait actions calculation. Since the mutation reduces the number of nodes or the number
of edges, the size of the graph reduces quickly towards the optimal conflict graph.
\extended{The graph reduction search algorithm stops when no further mutations yield performance improvements (Algorithm~\ref{alg:conflict_graph_training} line \ref{line:alg.graph_reduct.stop_condition_2}), or when all mutations of the current population have been exhausted (line \ref{line:alg.graph_reduct.stop_condition_1}).}
\maintext{{Our extended version~\cite{NeurCC-extended} provides more details of the graph reduction search, including the exploration of conflict graphs, the calculation of pipeline-wait actions, the computation of the function with optimized pipeline-wait actions, and the optimizations applied during pipeline waits.}}

The graph reduction search algorithm complements the Bayesian optimization for {\cacheFP}.  
{The former efficiently explores functions derived from conflict graphs, while the latter is effective at approximating high-performance but sparsely distributed functions.}
For instance, in the YCSB workload, we observe that a no-pipeline-wait policy can 
outperform learned pipeline-wait actions due to reduced computational costs associated with conflict detection and
dependency tracking.  Such a function cannot be identified through graph reduction search, but it can be found
quickly by Bayesian optimization within minutes.

\subsection{Function Optimization Pipeline}
\label{sec:training-pipeline}
Given a limited time budget, it is important to prioritize the optimization of actions that are more likely to yield significant
performance improvements. {\dbname} uses an optimization pipeline that transfers knowledge from one optimization stage
to the next. It allocates the time budget efficiently across these stages. Recall that the function {\cacheF} is
decomposed into three sub-functions {\cacheFP}, {\cacheFD}, and {\cacheFT} (see Section~\ref{sec:fine-tuning}).  The
optimization pipeline consists of four stages, each targeting a subset of actions while keeping the remaining actions
unchanged from the best-seen function. Each stage leverages the previously evaluated functions and their
corresponding performance scores as initial training data for the stage's optimizer.

The first stage optimizes both {\cacheFP} and {\cacheFD} using the graph reduction search algorithm. These two sub-functions are optimized first because they lead to the largest performance gains, e.g., with improvements of nearly 80\% in the TPC-C workload.
{\cacheFP} and {\cacheFD} can be optimized simultaneously, as the \textit{no detection} action and edge removal mutation both assume no conflicts around an operation.
This stage starts with the same initial configuration as that of IC3, in which all conflict detection actions are set to \textit{detect critical}. 
\maintext{The graph reduction search iteratively refines the conflict graph.
If a function input {{\inputstate}} maps to conflict graph nodes that are entirely disconnected from others, it indicates that database operations in state {{\inputstate}} rarely encounter conflicts. In such cases, we set the conflict detection action to \textit{no detection} to reduce transaction execution costs.}
\extended{{The graph reduction search iteratively refines the conflict graph.
When a function input {\inputstate} maps to conflict graph nodes that are entirely disconnected from others, it indicates that database operations in state {\inputstate} rarely encounter conflicts. In such cases, we set the conflict detection action to \textit{no detect} to reduce transaction execution costs (Algorithm~\ref{alg:get_updated_agent} lines \ref{line:alg.update_wait.isolated_node_start} -- \ref{line:alg.update_wait.isolated_node_end}).}}
We set $K = 4$ to enable quick convergence to an efficient function.
The stage ends when no further mutations yield performance improvements, typically within 10 minutes due to 
the monotonically reducing nature of the mutation process.
\maintext{
{The details on how we optimize {\cacheFP} and {\cacheFD} can be found in our extended version~\cite{NeurCC-extended}.}
}

The second stage optimizes {\cacheFT} using Bayesian optimization. At this stage, most parameters in {\cacheFT} are
continuous and exhibit good locality.  This stage ends when no further performance improvements are observed after
several consecutive evaluations (e.g., 20 evaluations).
The third stage is the same as the first stage but with $K = 8$, allowing for
deeper exploration of the conflict graph, in order to refine the conflict graph and functions.
Finally, the fourth stage fine-tunes all the parameters of {\cacheF} using Bayesian optimization. This stage runs until the time budget is exhausted.

\subsection{Feature Selector Optimization}
\label{sec:prune_opt}

{We define the feature selector {\encoder} as a function mapping a feature selection strategy to the system performance achieved with the selected features.}
Each feature selection strategy is represented as a vector, with each element indicating whether a corresponding feature is included and specifying its transformation type (linear, square root, or logarithmic).
The output of {\encoder} is a set of input features for  $\mathcal{F}$, which is subsequently used during the function optimization.
{As in Section~\ref{sec:bayesian_opt}, {\dbname} uses Bayesian optimization to iteratively propose new feature selection strategies, evaluate them, and refine the surrogate model.}

To balance effectiveness and efficiency, the feature selection strategy is optimized to maximize the performance improvement achieved through function optimization within a user-defined time budget.
Note that selecting too few features can overly constrain the search space of functions, leading to suboptimal solutions, whereas selecting too many features can expand the search space excessively, making the optimization computationally expensive.
Therefore, we balance the two factors by incorporating the time budget into the performance.
In our experiments, we run the feature selector optimization for one day, using a function optimization budget of one hour for each feature selection evaluation.

{The optimization is costly and is performed only once when the system is deployed. 
Subsequent changes in system load, contention, or access patterns do not automatically trigger re-optimization.
Instead, {\dbname} relies on function optimization to handle such workload drifts.
However, users can explicitly run the optimization to fine-tune the default selector for specific workloads.
Our feature selectors and their configurations are available in the code repository~\cite{NeurCC-impl}.}

\section{Deployment}
\label{sec:deployment}

{
To support practical adoption, we provide a complete guideline for deploying {\dbname} in a new environment. The process consists of the following four steps.

\mypar{Transaction Annotation (Optional)} When transactions are executed in the stored procedure mode, the user first assigns a \emph{transaction type} to each transaction and a unique \emph{access ID} to every SQL statement. The transaction type can be assigned per stored procedure, while the access ID can be derived from the statement order (e.g., the line number of the API invocation within the procedure~\cite{DBLP:conf/osdi/WangDWCW0021/polyjuice, DBLP:journals/pacmmod/NguyenCDA25/correct_assumption}). Next, the user constructs a large static conflict graph as the initial graph, where each node represents a type of transaction access (an access ID), and edges connect accesses on the same table if at least one is a write.

\mypar{Optimization Pipeline Setup} Next, the user configures the optimization pipeline for function optimization. This includes specifying the number of optimization steps, selecting the optimization algorithm for each step (e.g., graph reduction search or Bayesian optimization), defining termination conditions for Bayesian optimization steps, and setting the population size for graph reduction search. Our default optimization pipeline is described in Section~\ref{sec:training-pipeline}. The user may also select an initial function according to their needs. By default, {\dbname} initializes with {IC3} in stored procedure mode and {2PL} in interactive transaction mode.

\mypar{Feature Selector Optimization} After configuring the optimization pipeline, the user runs the feature selector optimization once to reduce the state space and accelerate future function optimization. As discussed in Section~\ref{sec:prune_opt}, this step runs for one day by default. Because it is costly, it is performed only once during initial deployment. Users may also choose to manually trigger this optimization periodically (e.g., once per month) or on demand (e.g., when migrating to a new machine).

\mypar{Execution and Function Optimization} Finally, the user runs {\dbname} with the optimized feature selector and the configured optimization pipeline. During execution, the system continuously monitors throughput and automatically detects workload drift (Section~\ref{sec:func_update}). When workload drift is detected, it automatically triggers function optimization to produce a better function for subsequent transaction processing.
}

\section{Performance Evaluation} \label{sec:evaluation}

In this section, we evaluate the performance of \dbname and compare it with state-of-the-art concurrency control (CC)
algorithms under diverse workloads.
We implement {\dbname} in C++ and integrate it into Silo~\cite{DBLP:conf/sosp/TuZKLM13/silo}, a widely used database system for evaluating CC algorithms.
Specifically, we add the following components to the database engine: a collector that captures transaction-related features, a feature selector that selects influential features, an in-database cache that stores the learned functions, and an optimizer
that tracks system performance for function optimization and workload drift detection. We integrate Algorithm~\ref{alg:bdta.exection} into the  concurrency control code of the database engine. We implement a script that processes optimization requests from the database.
The code is available at~\cite{NeurCC-impl}.
\maintext{Details on {\dbname} deployment are included in the extended version~\cite{NeurCC-extended}.}

\subsection{Experimental Setup}
\label{sec:evaluation.setup}

{All experiments are conducted on a 24-core 
 Intel(R) Xeon(R) Gold 5317 server with 128GB of memory. The CPU runs at a base frequency of 800 MHz, and a maximum turbo frequency of 3.6 GHz.}

\mypar{Execution modes} {\dbname} executes transactions in one of the two modes: stored procedure~\cite{DBLP:conf/sigmod/WangMCYCL16/ic3, DBLP:conf/osdi/WangDWCW0021/polyjuice, DBLP:conf/sigmod/SuCDAX17/Tebaldi,
DBLP:conf/icde/Su0Z21/C3, DBLP:conf/sigmod/GuoWYY21/bamboo, DBLP:conf/sigmod/LinC0OTW16/LEAP, DBLP:journals/vldb/YaoZLOX18/DGCC, DBLP:journals/pacmmod/NguyenCDA25/correct_assumption}, and interactive transaction~\cite{DBLP:journals/tocs/ZhangSSKP18/TAPIR, DBLP:journals/tocs/CorbettDEFFFGGHHHKKLLMMNQRRSSTWW13/spanner,
DBLP:conf/sigmod/PavloCZ12/skew_aware, DBLP:journals/pvldb/YuBPDS14/1000_core_cc}.
\begin{itemize}[leftmargin=*]
    \item \textbf{Stored procedure mode.} This mode represents how stored procedure transactions are executed by the
database engine. In particular, all the transaction's {statements} are known before execution. {As a result, concurrency
control algorithms can use transaction type and access ID as features. The former can be assigned to each stored procedure, and the latter can be derived from the statement order, e.g., the line number of the API invocation within the procedure~\cite{DBLP:conf/osdi/WangDWCW0021/polyjuice}.} The transaction only returns results to the
user after the transaction commits or aborts, which means it can read uncommitted data during execution and perform partial retries.

    \item \textbf{Interactive transaction mode.} This mode represents how interactive, user-driven transactions are
executed. In particular, the transaction's statements (SQL queries) are issued and
executed individually. The user waits for the response of each statement before issuing the next. Because the operations
are not known in advance, concurrency control algorithms cannot rely on features such as transaction type and access ID.
Furthermore, since the results are returned to the user immediately after each operation, the algorithms cannot perform partial retry and dirty read.
\end{itemize}

\mypar{Workloads}
{We conduct experiments on the widely used YCSB and TPC-C workloads. We change three types of workload settings for evaluation: access pattern, contention level, and system load.
The workloads and their settings are described below.}

\begin{itemize}[leftmargin=*]
    \item \textbf{YCSB-extended.}
    {We extend YCSB to group multiple read/write operations into a single transaction, and simulate varying access patterns by changing hotspot positions.}
    Specifically, we fix the list of read and write operations for all transactions, representing stable operations
found in real-world applications ~\cite{DBLP:conf/osdi/WangDWCW0021/polyjuice,
DBLP:conf/sigmod/WangMCYCL16/ic3}. {We then vary the positions of hotspots, where hotspot accesses follow a skewed
distribution (Zipfian factor $\theta = 1$), while other accesses follow a uniform distribution  across all keys (Zipfian factor $\theta =
0$).} For this workload, we use 16 threads.
    \item \textbf{TPC-C.} We use the standard TPC-C workload, which consists of five transaction types: two read-only transactions
    and three read-write transactions. This workload has complex inter-transaction
    dependencies, making it harder to optimize than YCSB-extended. We vary the contention level by changing the number of warehouses
    (fewer warehouses mean higher contention). We vary the system load by changing the number of threads (more threads mean a higher system load). For this workload, the default setting is 16 threads and one warehouse.
    {For a fair comparison, we follow {\polyjuice} to include all five types of TPC-C transactions. We learn the concurrency control actions for three types, i.e., \textit{payment}, \textit{new order}, and \textit{delivery}, while excluding the remaining two types, as they are read-only transactions.}
\end{itemize}

\maintext{{We have also conducted other experiments that evaluate {\dbname}'s function optimization time under interactive transaction mode, the impact of features, and the impact of transaction length.} Due to space constraints, we present these experiments in Section 7 of the extended version~\cite{NeurCC-extended}.}

\mypar{Baselines} We compare {\dbname} against five state-of-the-art concurrency control algorithms:
Silo~\cite{DBLP:conf/sosp/TuZKLM13/silo}, {\polyjuice}~\cite{DBLP:conf/osdi/WangDWCW0021/polyjuice},
IC3~\cite{DBLP:conf/sigmod/WangMCYCL16/ic3}, 2PL~\cite{2PL_no_wait}, and CormCC~\cite{DBLP:conf/usenix/TangE18/CormCC}.
For the first four baselines, we use the implementations in~\cite{DBLP:conf/osdi/WangDWCW0021/polyjuice}. For CormCC, we
follow the approach in {\polyjuice} and report the best performance between 2PL and Silo.
{We refer to this baseline as \textup{$CormCC^*$}.}

\subsection{Stored Procedure Mode}
\label{sec:evaluation.stored}

{{\dbname} and all the baselines support the stored procedure mode. We run all the experiments five times and report the average results. Each algorithm runs for three hours to match the time {\polyjuice} requires to reach optimal performance. We use IC3 as the initial function for function optimization.

\begin{figure}[ht]
  \centering
  \begin{subfigure}{0.49\linewidth}
    \includegraphics[width=\linewidth]{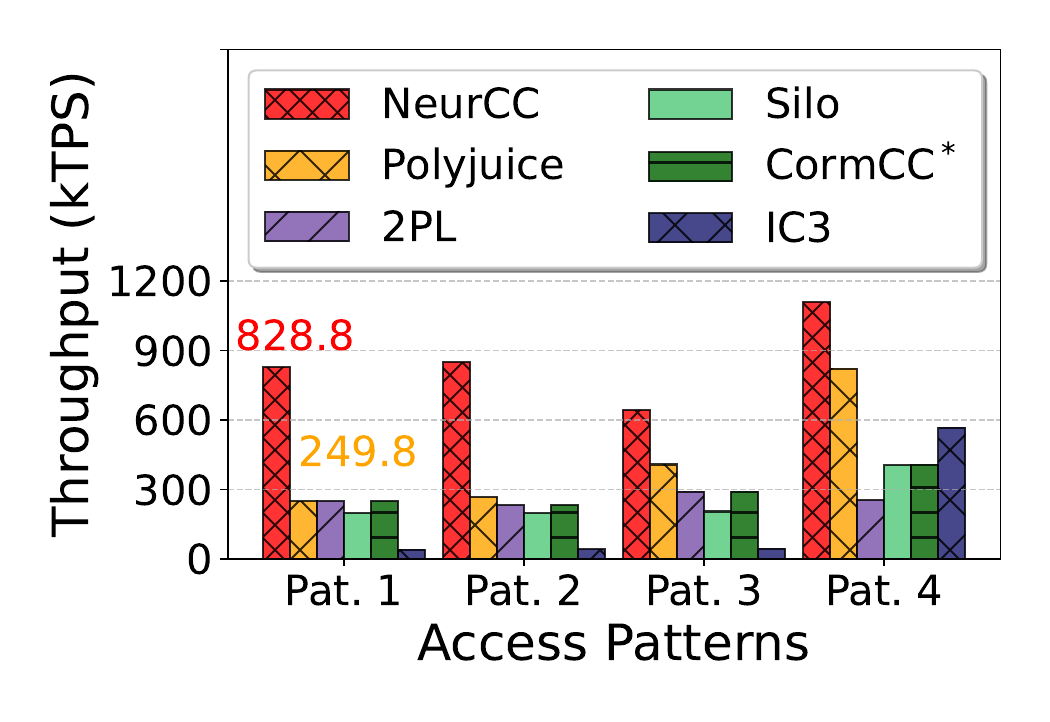}
    \vspace{-8mm}
    \caption{{Access pattern}}
    \label{fig:pattern}
  \end{subfigure}
  \begin{subfigure}{0.49\linewidth}
  \includegraphics[width=\linewidth]{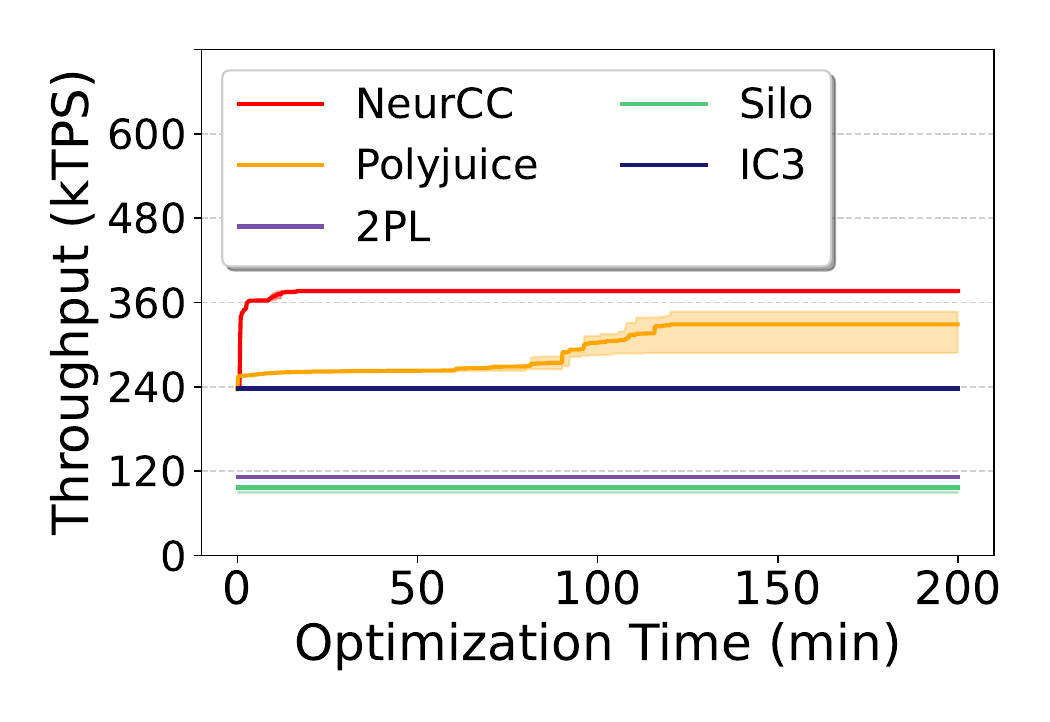}
    \vspace{-8mm}
    \caption{{High contention}}
    \label{fig:1wh}
  \end{subfigure}
  \begin{subfigure}  {0.49\linewidth}
  \includegraphics[width=\linewidth]{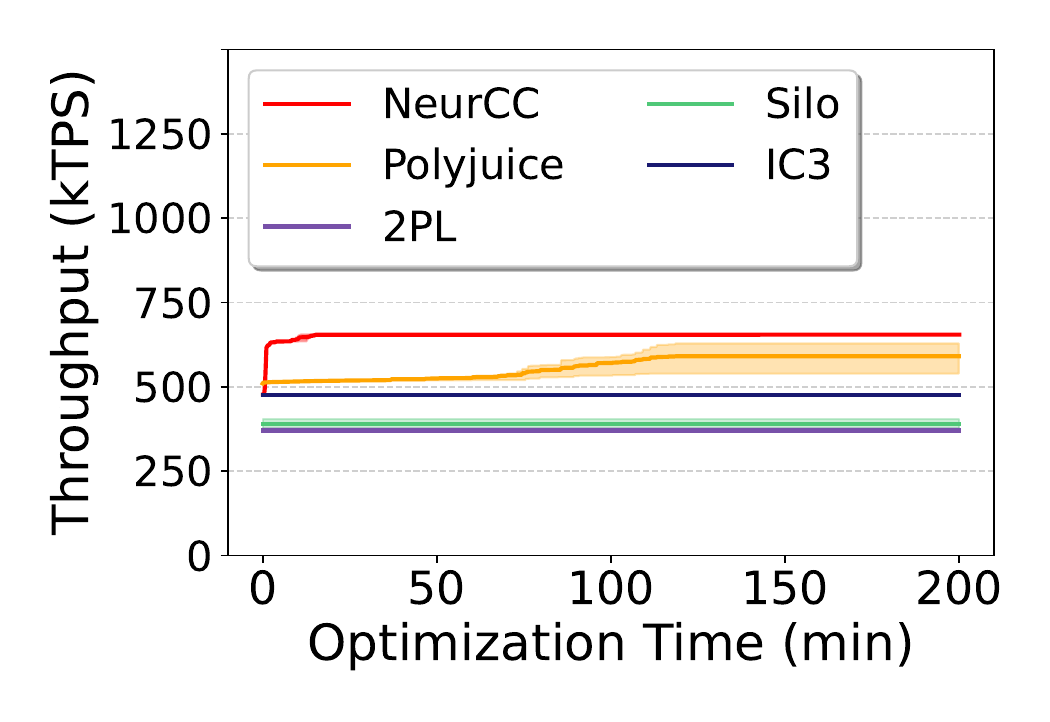}
    \vspace{-8mm}
    \caption{{Low contention}}
    \vspace{-3mm}
    \label{fig:4wh}
  \end{subfigure}
  \begin{subfigure}{0.49\linewidth}
  \includegraphics[width=\linewidth]{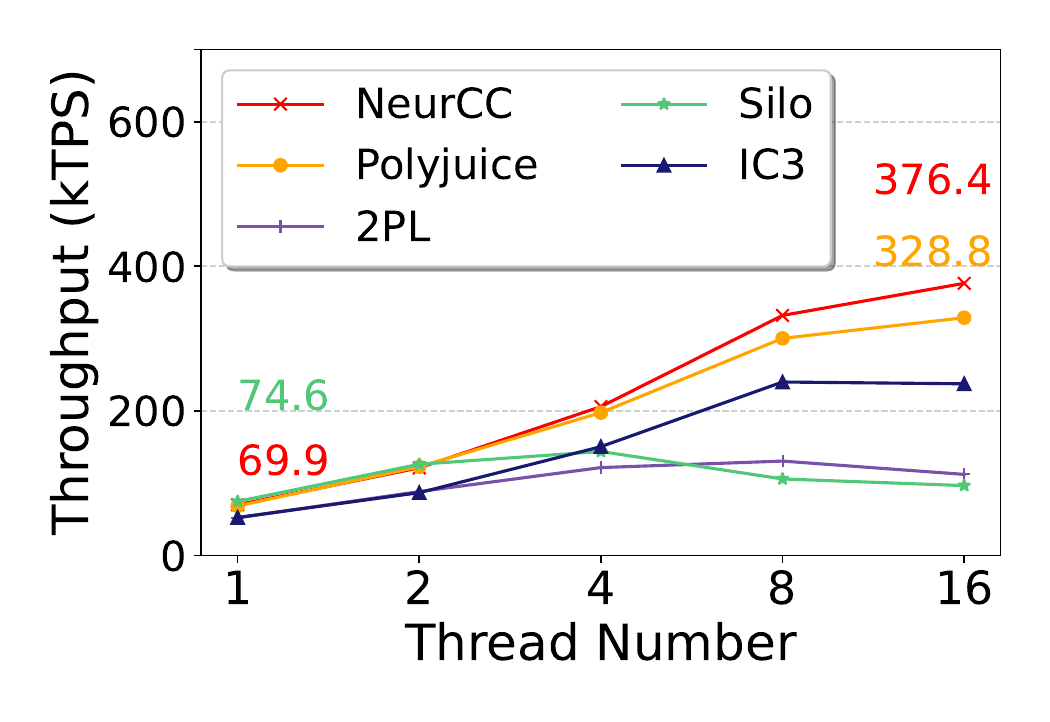}
    \vspace{-8mm}
    \caption{System load}
    \vspace{-3mm}
    \label{fig:scalability}
  \end{subfigure}
  \caption{Performance comparison between {\dbname} and baselines under the YCSB-extended and TPC-C workloads in the stored procedure mode.}
  \vspace{-1mm}
  \Description{Performance comparison between {\dbname} and baselines under the YCSB-extended and TPC-C workloads in the stored procedure mode.}
  \label{fig:stored}
\end{figure}

\mypar{Varying access pattern}
{We fix the contention level and system load while varying the access pattern using four different hotspot distributions}.\footnote{{The values of 1 and 0 indicate hotspot and uniform access, respectively.  The four access patterns, from the left to the right in Figure~\ref{fig:pattern}, are $(0,0,0,1,0,0,0,0,0,0)$, $(1,0,0,1,1,0,0,0,0,0)$, $(0, 0, 0, 1, 0, 1,0, 1,1,0)$, and $(1,1,0,0,1,0,0,1,1,0)$.}}
Figure~\ref{fig:pattern} shows the results for the YCSB-extended workload. The hotspot distributions, shown
from left to right, shift from being skewed at the beginning of the transaction to being skewed at the end.
We observe that {\dbname} has the best throughput, up to 3.32{$\times$}, 4.38{$\times$}, and 4.27{$\times$} higher than that of Polyjuice, 2PL, and Silo, respectively.
The reason is that {\dbname} learns a set of optimized pipeline-wait actions, timeouts, and conflict detection actions, enabling efficient transaction pipelining for hotspots while reducing the overhead for non-conflicting operations.

The performance gap between {\dbname} and the second-best baseline decreases as the hotspot
distribution shifts toward the end of the transaction. This is because the learned pipeline-wait actions, which contribute the most to the performance advantage of {\dbname}, become less effective as the hotspots are closer to the end of the transactions. In particular, pipeline-wait  actions allow dirty reads, but the advantage of
dirty reads over clean reads depends on the trade-off between the gain from reducing block time and the overheads from
cascading aborts and other issues~\cite{DBLP:conf/sigmod/GuoWYY21/bamboo, DBLP:journals/pvldb/FaleiroAH17/EWV}. As the hotspots shifting towards the end, the gain from reducing block time becomes smaller.

\mypar{Varying contention level} We fix the thread number to 16 and measure the performance with one warehouse, representing high contention (Figure~\ref{fig:1wh}), and with four warehouses, representing low contention (Figure~\ref{fig:4wh}).
{
The shaded areas represent the interquartile range, i.e., the 25th to 75th percentiles of throughput across multiple runs, illustrating the fluctuations of throughput in different CC algorithms.
{\dbname} achieves up to 1.14$\times$ throughput of the second-best baseline.
{\dbname} also exhibits a more stable learning process compared to {\polyjuice}. 
{\polyjuice} uses random genetic search, which leads to  high fluctuations, whereas {\dbname} employs a surrogate model to guide optimization, which achieves better stability throughout the optimization process.
}

\mypar{Varying system load}
We fix the number of warehouses to one and vary the system load by increasing the number of threadsfrom 1 to 16. Figure~\ref{fig:scalability} shows the results. 
There are three main observations.
{First, {\dbname} outperforms all the baselines under high load (more than four threads).
This is because our algorithm finds a better combination of CC actions that allows for higher concurrency.}
Second, both {\dbname} and {\polyjuice} perform slightly worse (less than 6.2\% drop in throughput) than Silo under low load (one thread), even though they converge to the same CC actions.
{This gap is also reported in {\polyjuice}~\cite{DBLP:conf/osdi/WangDWCW0021/polyjuice}.}
{The reason is that the learning processes of {\dbname} and {\polyjuice} incur additional overheads, which do not exist in the non-learned Silo system.}
\revisionhint{R2.O7}{Third, {\dbname} achieves higher throughput than the baselines even when the workload does not saturate the system (e.g., with 4 threads).
This is because  {\dbname} can still extract more concurrency from the workload than the baselines.
}

\begin{figure}[htbp]
  \centering
  {
  \begin{subfigure}{0.89\linewidth}
    \includegraphics[width=\linewidth]{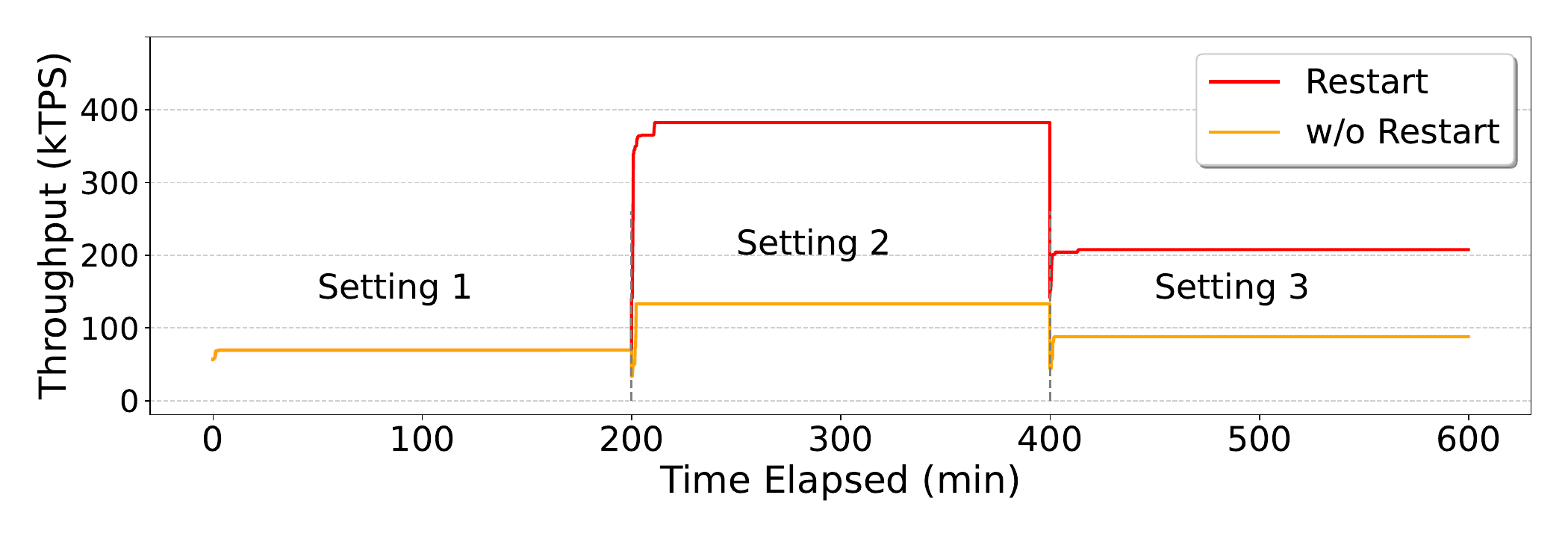}
   \vspace{-8mm}
    \caption{\textcolor{revR2}{Workload drift}}
    \label{fig:drift}
  \end{subfigure}
  }
  {
  \begin{subfigure}{0.49\linewidth}
    \includegraphics[width=\linewidth]{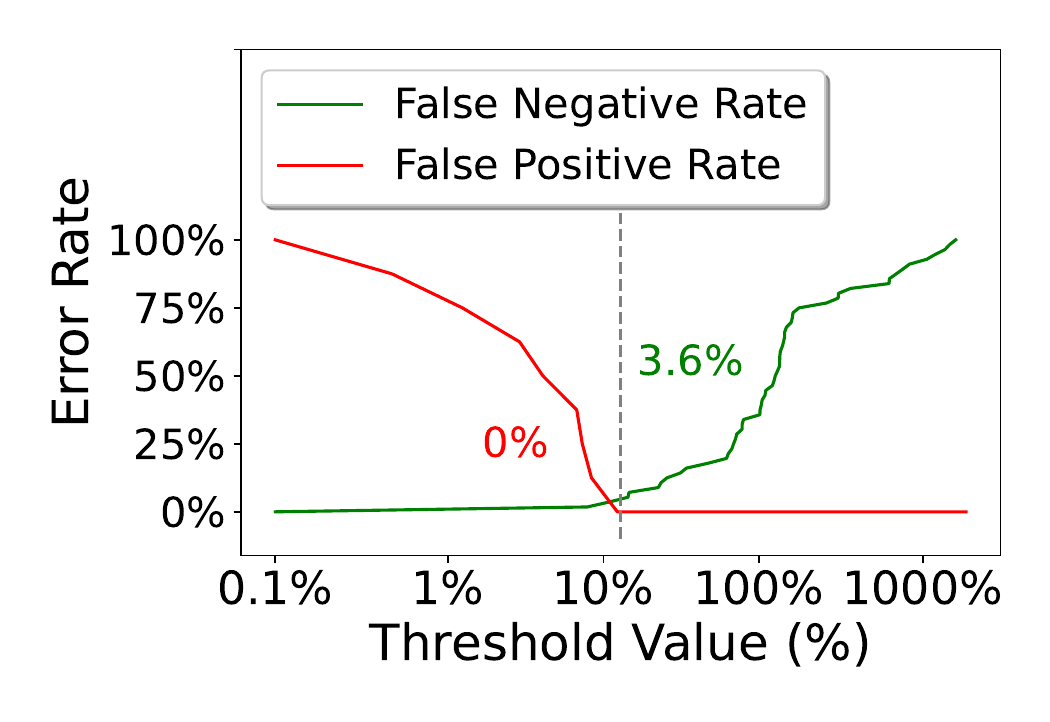}
   \vspace{-8mm}
    \caption{{Impact of threshold value}}
    \label{fig:accuracy}
  \end{subfigure}}
  \extended{
  \begin{subfigure}{0.49\linewidth}
    \includegraphics[width=\linewidth]{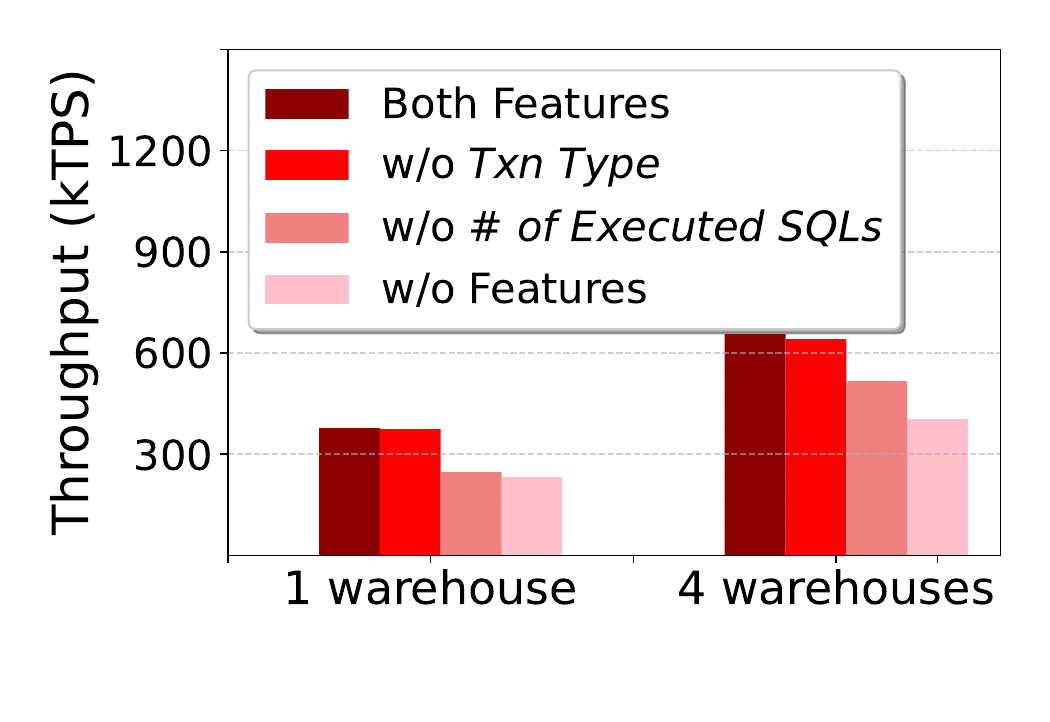}
    \vspace{-8mm}
    \caption{{Impact of features}}
    \label{fig:feature}
  \end{subfigure}}
    \begin{subfigure}  {0.49\linewidth}\includegraphics[width=\linewidth]{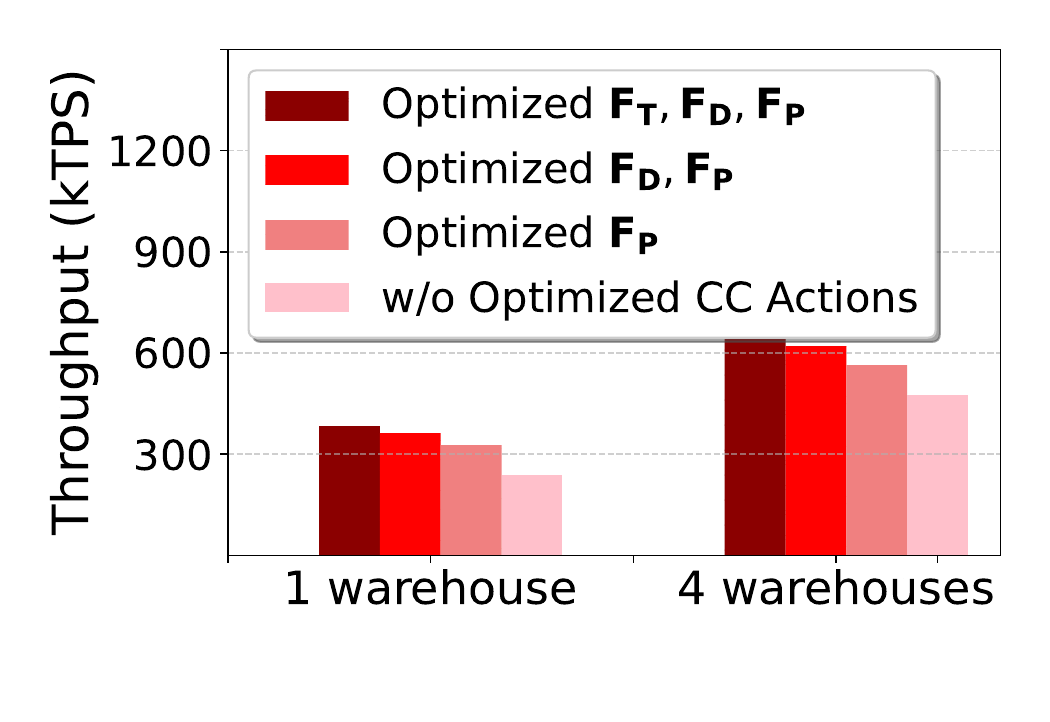}
    \vspace{-8mm}
    \caption{Impact of learned actions}
    \label{fig:factor-stored}
  \end{subfigure}
    \begin{subfigure}{0.49\linewidth}
    \includegraphics[width=\linewidth]{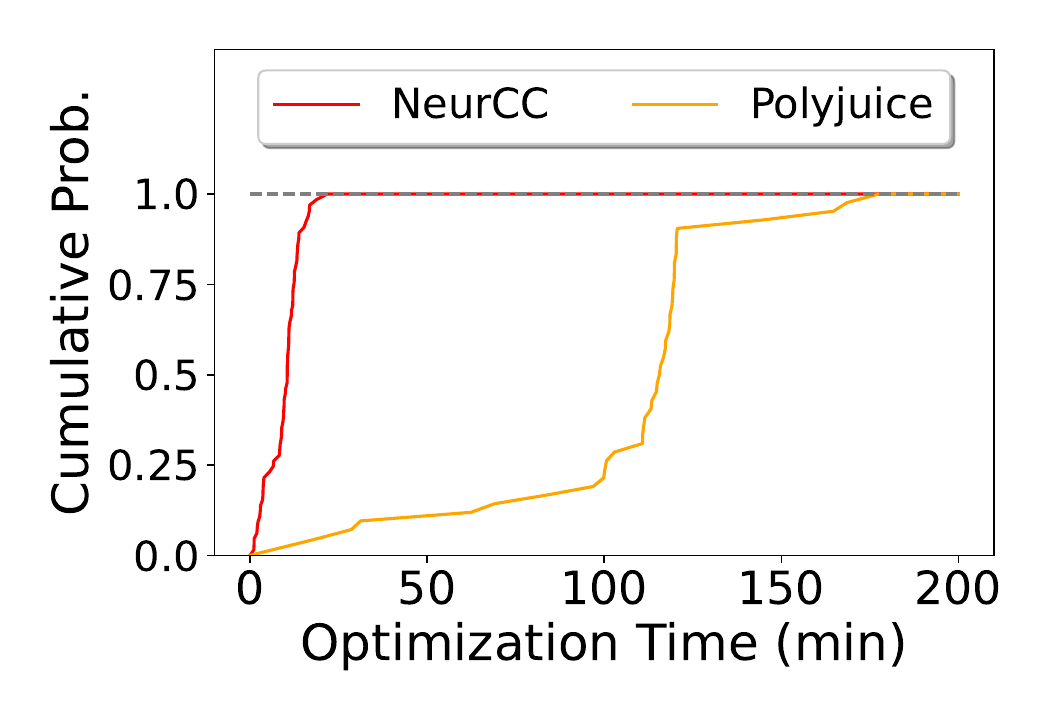}
    \vspace{-8mm}
    \caption{{Time to peak performance}}
    \maintext{\vspace{-3mm}}
    \label{fig:opt-time}
  \end{subfigure}
\begin{subfigure}{0.49\linewidth}
    \includegraphics[width=\linewidth]{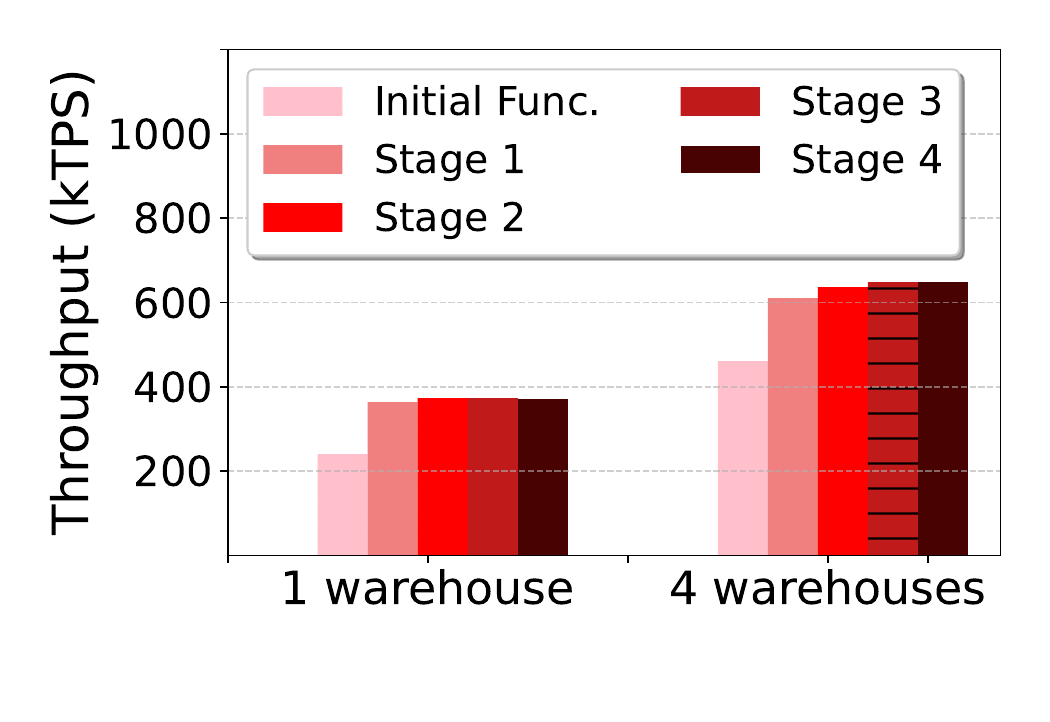}
    \vspace{-8mm}
    \caption{\textcolor{revR2}{Impact of optimization stages}}
    \vspace{-3mm}
    \label{fig:stages}
  \end{subfigure}

  \caption{Analysis of {\dbname} in the stored procedure mode.}
  \vspace{-2mm}
  \Description{Analysis of {\dbname} in the stored procedure mode.}
  \label{fig:stored-analysis}
\end{figure}

\mypar{Workload drift}
\revisionhint{R2.O7}{We simulate workload drifts under the TPC-C workload by changing the number of threads at different time intervals, i.e., 1 thread for Setting 1 (starting at minute 0), 16 threads for Setting 2 (starting at minute 200), and 4 threads for Setting 3 (starting at minute 400).
Figure~\ref{fig:drift} compares the system throughput when optimizing from the initial function (``Restart''), versus when optimizing from the previously learned function (``w/o Restart'').
It can be seen that {\dbname} reacts quickly to workload drifts.
Furthermore, the ``w/o Restart'' approach fails to achieve optimal performance within three hours in both Setting 2 and Setting 3.
This is because our graph reduction search algorithm  involves only edge removal and node merging mutations. We note that in order to fully explore the conflict graph search space, the optimization process must begin with the full conflict graph, containing all nodes and edges.
{Therefore, starting the optimization from a previously learned optimal function, where the conflict graph has already been reduced, restricts the exploration of search space, and consequently leads to a suboptimal conflict graph for the new workload.}}

{
\mypar{Impact of threshold value}
We use eight different TPC-C workload settings\footnote{\label{note:tpcc}{Denote $(x,y)$ as $x$ threads and $y$ warehouses, the eight settings are $(1,1)$, $(2,1)$, $(4,1)$, $(8,1)$, $(16, 1)$, $(16, 2)$, $(16, 4)$, and $(16, 8)$.}} and evaluate how the threshold value affects the accuracy of drift detection when switching the workload among these settings.
We use false positive rate and false negative rate as the accuracy metrics. For the false positive rate, we set the workload to one of the eight settings and report the percentage of the experiment runs where a workload drift is falsely detected despite no actual workload drift occurring.
For the false negative rate, we evaluate all $7 \times 8$ possible workload drifts, starting from one of the eight settings and changing to another, and report the percentage of the experiment runs where these drifts are not detected.
As shown in Figure~\ref{fig:accuracy}, the threshold value of 10\% leads to robust detection (3.6\% false negative rate) without over-triggering the optimization process (0\% false positive rate).}

\extended{\mypar{Impact of features} 
{We conduct an ablation study to evaluate the impact of the two features in the optimized feature set for the TPC-C workload. These features are the transaction type and the number of executed SQLs. Figure~\ref{fig:feature} shows the results for settings with one and four warehouses, in which  features are progressively removed from the optimized feature set. It can be seen that both features are necessary for high performance.}}

\mypar{Impact of learned actions}
{We evaluate the impact of the learned actions by reverting the learned CC actions in the final learned function to the actions in the initial function (IC3).
Figure~\ref{fig:factor-stored} shows that  different actions have different impacts.
Among them, the {\cacheFD} and {\cacheFP} sub-functions  (see Section~\ref{sec:fine-tuning}) --- ones that are prioritized in our optimization pipeline --- account for the majority of the performance improvement (85.8\% and 80.6\% in the one warehouse and 4 warehouse settings, respectively).
The reason is that, although dirty reads allow for a higher degree of concurrency between transactions, they also introduce the risk of dependency cycles that can severely degrade performance.
By optimizing {\cacheFD} and {\cacheFP}, {\dbname}  mitigates this risk while ensuring high concurrency.
}

\mypar{Impact of optimization stages}
\revisionhint{R2.O6}{We evaluate how each optimization stage contributes to the agent function. Starting from the initial agent function, we measure the system throughput after each optimization stage. Figure~\ref{fig:stages} shows the results under TPC-C workloads with one and four warehouses.
It can be seen that the highest improvements are achieved in the early stages. In particular, the largest performance gains are observed in Stage 1, where the optimization of {\cacheFD} and {\cacheFP} are prioritized. The later stages yield marginal improvements, indicating that the agent function has likely converged. The results  demonstrate that the  proposed optimization pipeline is effective: by prioritizing influential actions early, {\dbname} achieves fast convergence, which is useful given the limited optimization time budgets.}

\mypar{Optimization time}
{Figure~\ref{fig:opt-time} compares the CDF of the optimization time for {\dbname} and {\polyjuice} to reach peak performance. 
It can be seen that {\dbname} has a significantly lower average optimization time (9.67 minutes) compared to {\polyjuice} (1.78 hours), or an 11{$\times$} speedup. Further, the optimization time of {\dbname} exhibits a lower variance than that of {\polyjuice}, because it uses graph reduction search guided by an online-trained surrogate model, whereas {\polyjuice} uses a random genetic search algorithm.}

\subsection{Interactive Transaction Mode}
\label{sec:evaluation.interactive}

{We compare {\dbname} against three other baselines that support the interactive transaction mode, namely 2PL, Silo, and CormCC.}
{In this mode, we use 2PL as the initial function for function optimization.}

\mypar{Varying contention level} Figure~\ref{fig:1wh-interactive} and Figure~\ref{fig:4wh-interactive} show the performance under one warehouse, i.e. high contention, and four warehouses, i.e. low contention.
{{\dbname} outperforms the baselines under both settings, achieving up to 1.96${\times}$ higher throughput than that of the second-best baseline.}
This gap is due to {\dbname}'s learned conflict detection actions and wait priorities.
For example, under the one warehouse setting, it learns to perform early validation after accessing hotspots (e.g.,
before a \textit{new order} transaction reads the ITEM table), which leads to a timely abort when the transaction is likely to
fail. It also assigns a low priority to early operations on the DISTRICT table, which reduces the likelihood of
unresolvable conflicts. It ensures that a \textit{new order}/\textit{payment} transaction $T_1$, which has completed a
\textit{read-modify-write} operation on the DISTRICT table, is ordered before another \textit{new order}/\textit{payment} transaction $T_2$
that has not yet performed such operations. This way, $T_2$ does not miss updates from $T_1$, allowing them to commit
together.

\begin{figure}[ht]
  \centering
  \begin{subfigure}  {0.49\linewidth}\includegraphics[width=\linewidth]{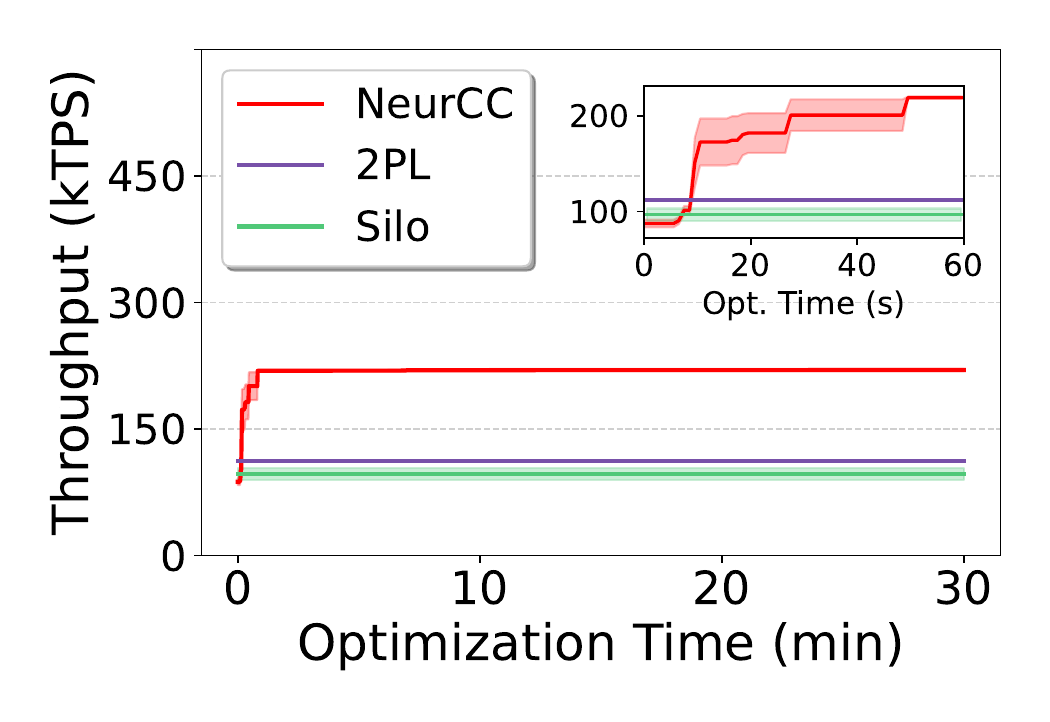}
    \vspace{-8mm}
    \caption{{High contention}}
    \label{fig:1wh-interactive}
  \end{subfigure}
  \begin{subfigure}  {0.49\linewidth}\includegraphics[width=\linewidth]{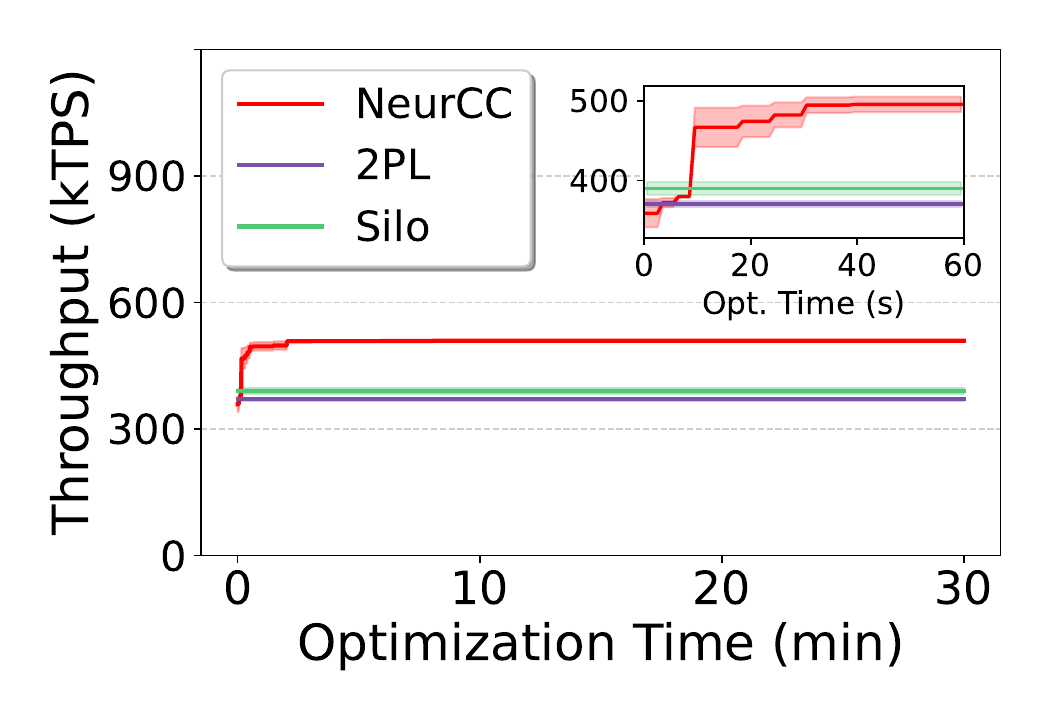}
    \vspace{-8mm}
    \caption{{Low contention}}
    \label{fig:4wh-interactive}
  \end{subfigure}
  \begin{subfigure}{0.49\linewidth}
    \includegraphics[width=\linewidth]{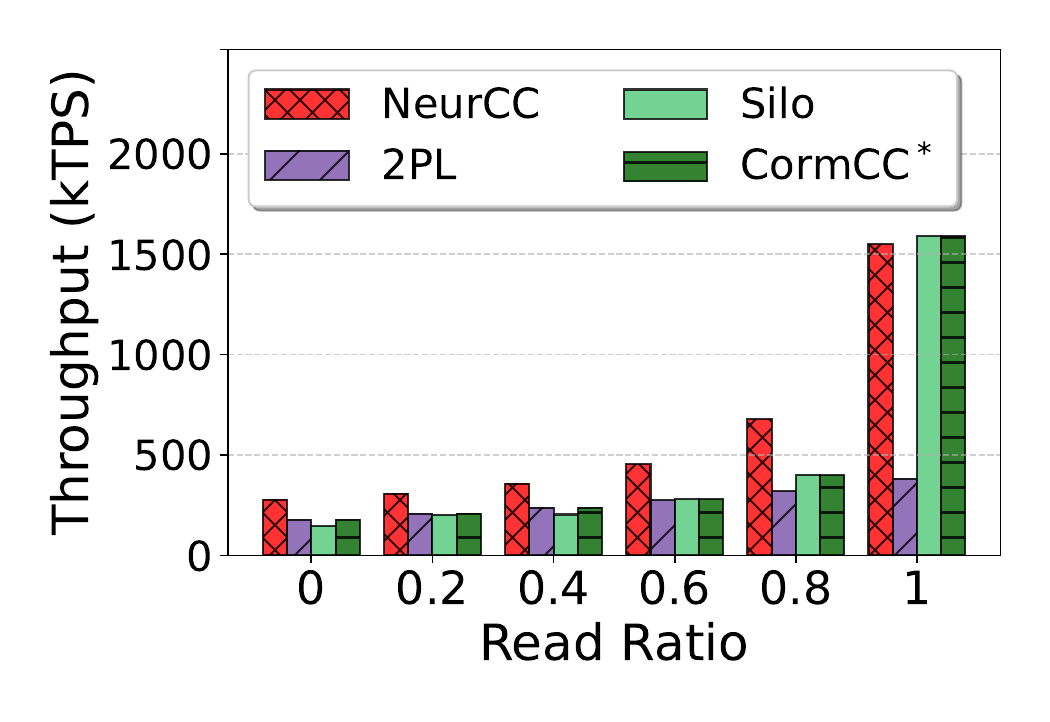}
    \vspace{-8mm}
    \caption{{Read ratio}}
    \maintext{\vspace{-3mm}}
    \label{fig:read-ratio}
  \end{subfigure}
  \extended{\begin{subfigure}{0.49\linewidth}
    \includegraphics[width=\linewidth]{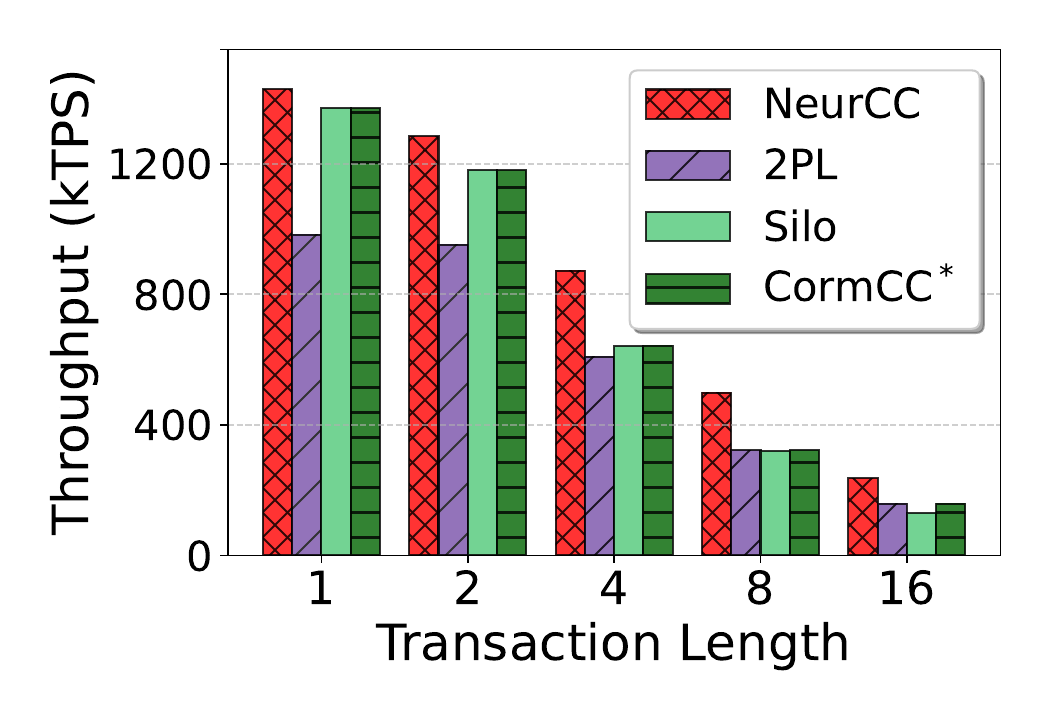}
    \vspace{-8mm}
    \caption{{Transaction length}}
    \label{fig:txn-length}
  \end{subfigure}}
  \begin{subfigure}{0.49\linewidth}
    \includegraphics[width=\linewidth]{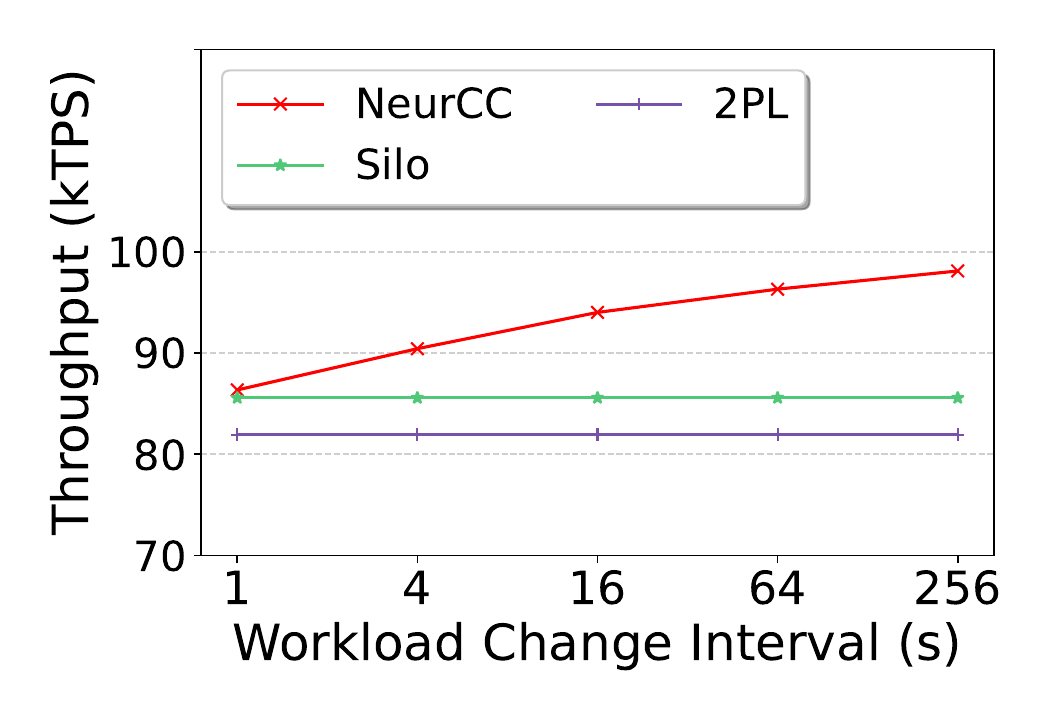}
    \vspace{-8mm}
    \caption{\textcolor{revR1}{Stress test}}
    \vspace{-3mm}
    \label{fig:stress}
  \end{subfigure}
  \caption{Performance comparison between {\dbname} and baselines in the interactive transaction mode.}
  \Description{Performance comparison between {\dbname} and baselines in the interactive transaction mode.}
    \vspace{-4mm}
  \label{fig:interactive}
\end{figure}

\maintext{\mypar{Varying read ratio} {To understand the impacts of the read ratio, we perform one set of experiments using the YCSB-extended workload. Specifically, we fix the transaction length at 10 and vary the read ratio (i.e., the percentage of read operations) from 0\% to 100\%.
The results are shown in Figure~\ref{fig:read-ratio}.
The throughput of all algorithms increases as the read ratio increases.
This trend is expected because higher read ratios result in fewer conflicts among transactions.
Further, {\dbname} outperforms the second-best baselines by up to 1.7{$\times$} when the read ratio is below 100\%. For read-only workloads, {\dbname} is slightly worse (by 2.5\%) than Silo. 
The reason is that learning-based algorithms can learn complex actions that deliver high performance when there are many conflicts, but they are less effective when there are few conflicts, due to the inherent overhead in the learning process.
}}

\extended{
\mypar{Varying read ratio and transaction length} {To understand the impacts of the read ratio and transaction length, we perform two sets of experiments using the YCSB-extended workload. First, we fix the transaction length at 10 and vary the read ratio (i.e., the percentage of read operations) from 0\% to 100\%. Second, we fix the read ratio at 50\% and vary the transaction length from 1 to 16.
The results are shown in Figure~\ref{fig:read-ratio} and Figure~\ref{fig:txn-length}.
The throughput of all algorithms increases as the read ratio increases or the transaction length decreases.
This trend is expected because higher read ratios or fewer transaction operations result in fewer conflicts among transactions.
Further, {\dbname} outperforms the second-best baselines by up to 1.7{$\times$} when the read ratio is below 100\%. For read-only workloads, {\dbname} is slightly worse (by 2.5\%) than Silo. 
The reason is that learning-based algorithms can learn complex actions that deliver high performance when there are many conflicts, but they are less effective when there are few conflicts, due to the inherent overhead in the learning process.
}}

\mypar{Rapid workload changes}
\revisionhint{R1.O3}{
Although real-world workloads typically evolve gradually~\cite{DBLP:conf/sigmod/MaAHMPG18/forecasting, DBLP:conf/sigmod/DasLNK16/demand-driven}, rapid workload changes can still arise from planned events such as
yearly promotional and sales events.
To evaluate {\dbname}'s robustness under such rapid changes, we vary the number of client threads from 1 to 16 at different switching frequencies. Figure~\ref{fig:stress} shows the average throughput over 1 hour. It can be seen that {\dbname} performs consistently better than the baselines, even under highly unstable workloads.
At the highest switching frequency, i.e. 1 second, {\dbname} achieves 85.90 ktps, which is comparable to the best static baseline (Silo achieves 85.57 ktps). The results demonstrate that {\dbname} is robust under rapid workload changes.}

\extended{

\begin{figure}[ht]
  \centering
  \includegraphics[width=0.49\linewidth]{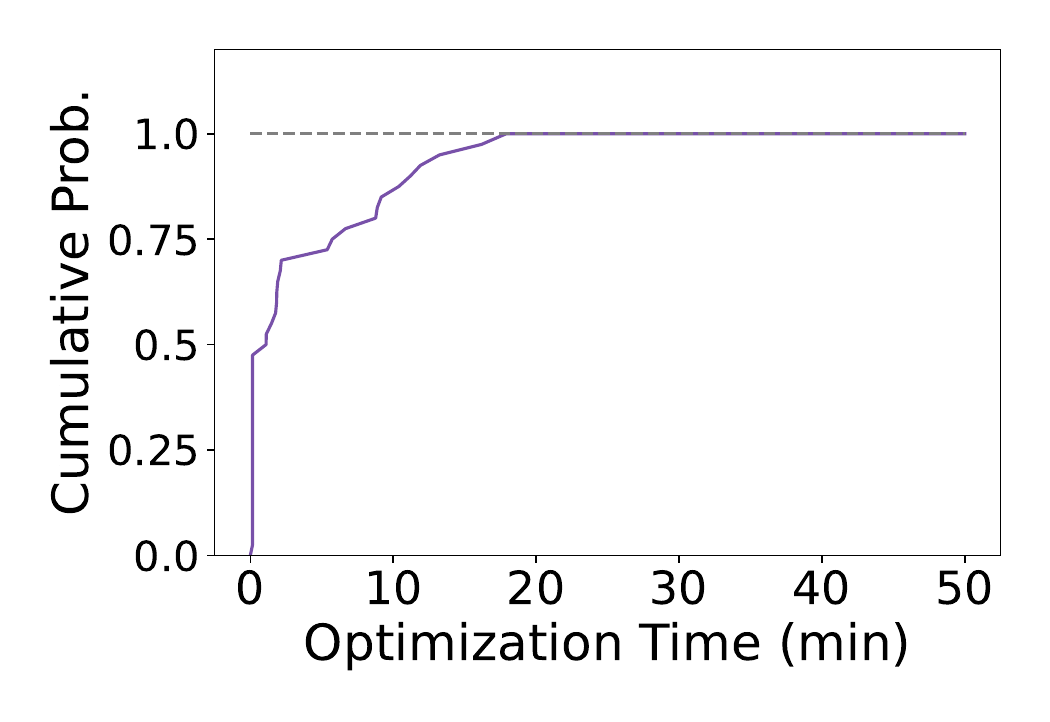}
  \vspace{-4mm}
  \caption{Time to peak performance in the interactive transaction mode.}
  \label{fig:opt-time-inter}
  \Description{Time to peak performance of {\dbname} in the interactive transaction mode.}
\end{figure}

\mypar{Optimization time}
{Figure~\ref{fig:opt-time-inter} depicts the cumulative distribution function of the optimization time for {\dbname} to reach peak performance. 
Compared to the stored procedure mode (Figure~\ref{fig:opt-time}), the optimization time in the interactive transaction mode is lower (3.59 minutes on average)
because {\dbname} does not need to learn $\cacheFP$
for the fact that
dirty reads are not permitted in this mode.}}

\begin{figure}[ht]
  \centering
  \begin{subfigure}{0.93\linewidth}
  \includegraphics[width=\linewidth]{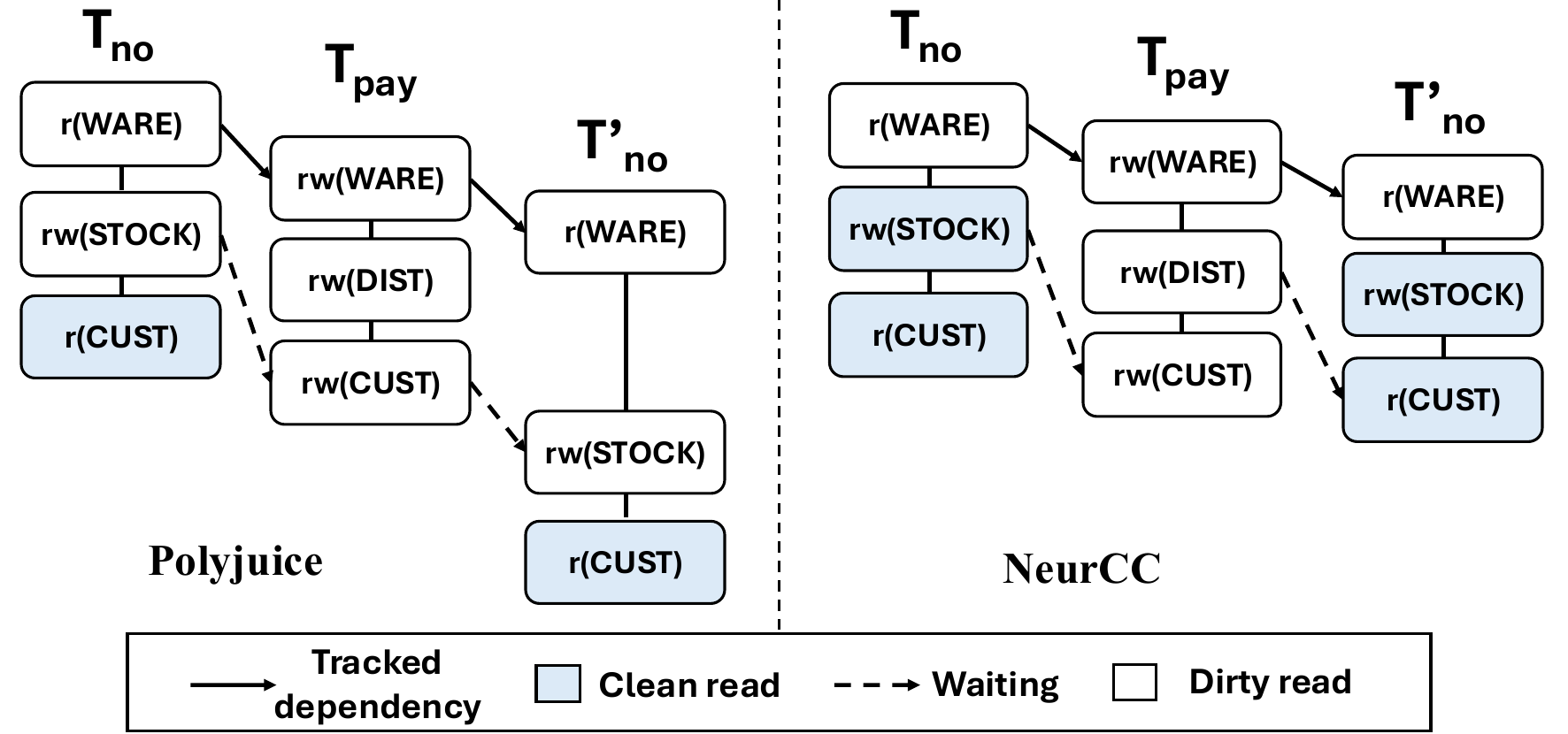}
  \caption{{Stored procedure mode}}
    \label{fig:case-stored}
  \end{subfigure}
  \begin{subfigure}{0.88\linewidth}
  \includegraphics[width=\linewidth]{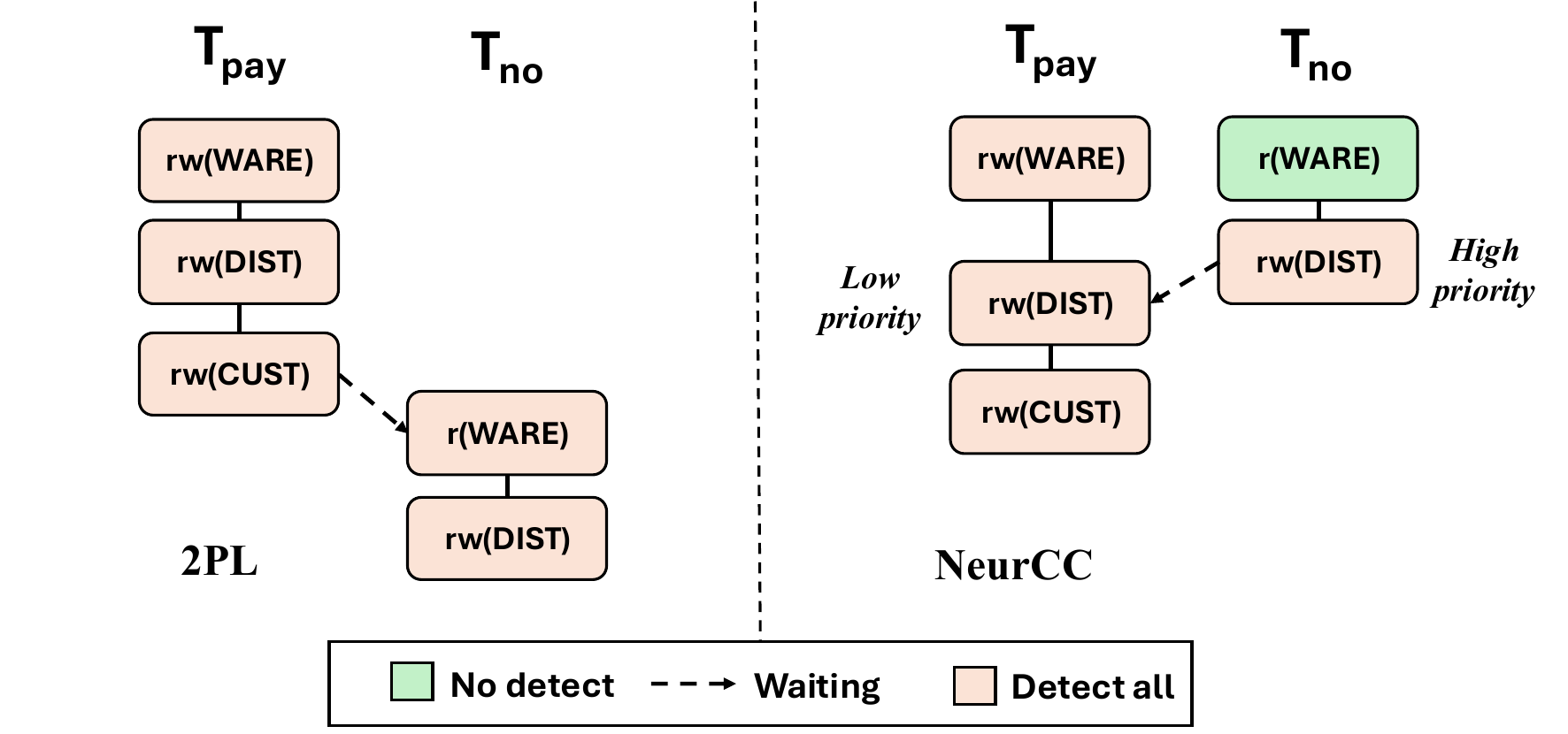}
  \caption{{Interactive transaction mode}}
    \label{fig:case-inter}
  \end{subfigure}
  \extended{
  \begin{subfigure}{0.55\linewidth}
  \includegraphics[width=\linewidth]{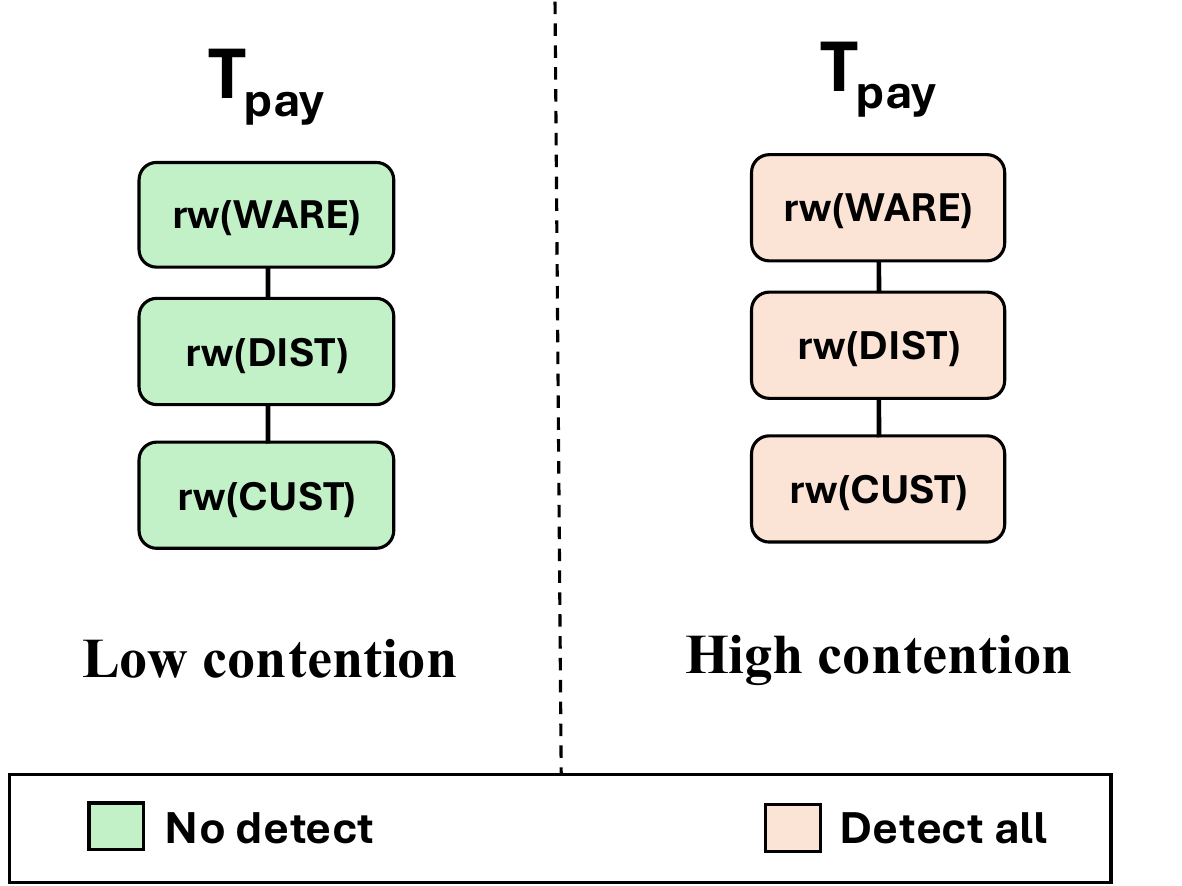}
  \caption{{Low vs.\ high contention}}
    \label{fig:case-contention}
  \end{subfigure}}
  %
  \caption{{Examples of the  learned actions.}}
  \Description{Case studies for learned actions.}
    \label{fig:case}
\end{figure}

{
\subsection{Examples of Learned Actions}
\label{sec:evaluation.cases}

Figure~\ref{fig:case-stored} and Figure~\ref{fig:case-inter} show examples that compare the CC actions learned by {\dbname} with those learned by the second-best baselines.
\revisionhint{R2.O4}{
While our implementation uses row-level locking, we show only table names in the examples for clarity.
}

In the stored procedure mode, {\polyjuice} is the second-best baseline.  As shown in Figure~\ref{fig:case-stored}, both {\dbname} and {\polyjuice} read dirty data during WAREHOUSE table accesses and track the transaction dependency $T_{no} \to T_{pay} \to T'_{no}$.
{\polyjuice} learns to perform the STOCK table access of $T'_{no}$ only after the STOCK table access of $T_{no}$, preventing a potential dependency cycle $T_{no} \to T_{pay} \to T'_{no} \to T_{no}$. 
In contrast, {\dbname} learns a conflict graph where STOCK table accesses of $T_{no}$ and $T'_{no}$ are unlikely to conflict with each other.
Therefore, it enables a more efficient transaction interleaving where the STOCK table access from $T_{no}$, the DISTRICT table access from $T_{pay}$, and the STOCK table access from $T'_{no}$ can be executed in parallel.

In the interactive transaction mode (Figure~\ref{fig:case-inter}), 2PL is the second-best basline, and it resolves the conflicts between $T_{no}$ and $T_{pay}$ by performing the WAREHOUSE table access of $T_{no}$ only after $T_{pay}$ commits. This results in a long execution time. On the other hand, {\dbname} learns to order $T_{no}$ before $T_{pay}$, allowing $T_{no}$ to optimistically access the WAREHOUSE table without waiting for $T_{pay}$ to commit.
It also learns to assign a high priority to the DISTRICT table access of $T_{no}$, ordering this access before that of $T_{pay}$. This prevents a dependency cycle $T_{no} \to T_{pay} \to T_{no}$. The safe interleaving of the two transactions allows them to commit faster in {\dbname}.  
}
\revisionhint{R2.O1}{This example demonstrates that by enabling better interleaving of transactions, {\dbname} can outperform 2PL despite incurring extra validation costs.}

\extended{
{
Figure~\ref{fig:case-contention} compares the CC actions learned by {\dbname} under two different contention levels. The system runs in interactive transaction mode, and we present the learned actions for both low contention (1 thread) and high contention (16 threads) settings. The optimal function varies significantly across workloads: under low contention, an OCC-like function performs best because of its low conflict detection overhead, whereas under high contention, a function that detects conflicts and resolves them immediately achieves superior performance.
}
}

\section{Related Work}
\label{sec:related}

Adaptive CC algorithms aim to achieve high performance across different workloads. They follow a
classify-then-assign workflow: analyzing workload information, classifying transactions based on manually selected
features, and assigning the most suitable CC algorithm for each type of transaction.  
The assignment step can be either heuristic~\cite{DBLP:conf/icde/Su0Z21/C3, DBLP:conf/mascots/TaiM96/AEP,
DBLP:conf/icde/LinCZ18/ASOCC, DBLP:conf/usenix/TangE18/CormCC, DBLP:conf/sigmod/SuCDAX17/Tebaldi, DBLP:journals/pacmmod/HuangZZXW25/PreemptDB, DBLP:journals/pvldb/HongZLDCPZ25/HDCC} or
based on learning~\cite{DBLP:conf/cidr/TangJE17/ACC, DBLP:conf/osdi/WangDWCW0021/polyjuice, DBLP:journals/pvldb/ZhuangLLCSZSPD25/TxnSails}. 

Among the algorithms adopting heuristic assignment, C3~\cite{DBLP:conf/icde/Su0Z21/C3} uses locality-sensitive
hashing (LSH) to cluster transactions with similar SQL texts, which tend to conflict on shared data objects.  It applies
2PL for intra-group conflicts and OCC for inter-group conflicts. AEP~\cite{DBLP:conf/mascots/TaiM96/AEP} predicts
whether a distributed transaction will encounter conflicts based on local data, then uses EWL when there are no
conflicts, and uses PSL when there are. MCC~\cite{DBLP:conf/sosp/XieSLAK015/Callas} and
Tebaldi~\cite{DBLP:conf/sigmod/SuCDAX17/Tebaldi} classify transactions based on prior workload knowledge, constructing
an algorithm hierarchy that switches between RP, SSI, and 2PL, depending on conflict type.
ASOCC~\cite{DBLP:conf/icde/LinCZ18/ASOCC} categorizes data into hot, warm, and cold based on the contention level, then
uses OCC for cold data, 2PL for hot data, and early validation with retry for warm data.  {\dbname} covers more CC
designs than these algorithms, and it learns a fine-grained and flexible combination of them for better performance.

{{\dbname} shares similarities with other algorithms that adopt learning-based assignment.}
{\polyjuice}~\cite{DBLP:conf/osdi/WangDWCW0021/polyjuice} classifies transaction operations by their access ID (a
hard-coded unique transaction access identifier) and  uses an evolutionary algorithm to optimize pipeline-wait actions
and transaction back-offs. ACC~\cite{DBLP:conf/cidr/TangJE17/ACC} clusters transactions similar to PartCC~\cite{DBLP:journals/pvldb/KallmanKNPRZJMSZHA08/HStore} and assigns
the best algorithm among OCC, 2PL, and PartCC to each cluster. For fast reconfiguration, it also trains a 
model that predicts the optimal algorithm for each cluster, enabling automatic online reconfiguration with
offline-trained models. {\dbname} supports more CC design choices than these algorithms, and achieves better
performance across more diverse workloads. Furthermore, it does not require prior knowledge of the workloads for access
ID assignment or transaction clustering.

\section{Conclusions} \label{sec:conclusion}

In this paper, we presented {\dbname}, a novel learned concurrency control algorithm that achieves high performance across diverse workloads. {\dbname} models concurrency control as a learnable function. It learns design choices from a large number of existing concurrency control algorithms. The algorithm employs an optimization algorithm that quickly converges to a combination of concurrency control actions that achieves high transaction throughput. Specifically, it uses Bayesian optimization and graph reduction search to reduce the overhead of performance  evaluations, thereby speeding up the optimization process.
Our evaluation shows that {\dbname} delivers high throughput, outperforming the baselines. In particular, its throughput is up to 3.32${\times}$, 4.38${\times}$, and 4.27${\times}$ higher than that 
of {\polyjuice}, 2PL, and Silo, respectively. 
Further, {\dbname}'s optimization time is up to 11$\times$ faster than that of other learned baselines, making  it more robust and adaptive to workload drifts.

\balance


\bibliographystyle{ACM-Reference-Format}
\bibliography{main-bibliography}

\end{document}